\documentclass[11pt]{article}
\usepackage[letterpaper,margin=1in]{geometry}
\usepackage[T1]{fontenc}
\usepackage{amsmath,amsthm}
\usepackage{newtxtext,newtxmath}
\usepackage{bm}
\usepackage{booktabs,threeparttable,graphicx,caption,array}
\usepackage{tikz}
\usetikzlibrary{circuits.logic.US,calc,decorations.pathreplacing}
\usepackage[authoryear,round]{natbib}
\usepackage{setspace}
\usepackage{microtype}
\usepackage[hidelinks,hypertexnames=false]{hyperref}
\usepackage[capitalise,noabbrev]{cleveref}
\usepackage{enumitem}
\usepackage{placeins}
\crefname{assumption}{Assumption}{Assumptions}
\crefname{example}{Example}{Examples}
\crefname{definition}{Definition}{Definitions}

\newtheorem{proposition}{Proposition}
\newtheorem{corollary}{Corollary}
\newtheorem{lemma}{Lemma}
\newtheorem{assumption}{Assumption}
\theoremstyle{definition}
\newtheorem{definition}{Definition}
\newtheorem{example}{Example}

\newcommand{\E}{\mathbb{E}}
\newcommand{\1}{\mathbf{1}}
\newcommand{\x}{\mathbf{x}}
\newcommand{\U}{\mathcal{U}}
\newcommand{\F}{\mathcal{F}}
\DeclareMathOperator{\spn}{span}

\title{The Logic of Factor Portfolios:\\ Boolean Rules, Interactions, and Trading Costs}
\author{Ziyao Wang\thanks{Department of Mathematics and Statistics, Texas Tech University. Corresponding author: \href{mailto:ziywang@ttu.edu}{ziywang@ttu.edu}.}
\and Svetlozar T.\ Rachev\thanks{Department of Mathematics and Statistics, Texas Tech University.}
\and Frank J.\ Fabozzi\thanks{Carey Business School, Johns Hopkins University.}}
\date{November 2026}

\begin{document}
\maketitle
\thispagestyle{empty}

\begin{abstract}
\noindent Investors turn signals into holdings through decision rules: buy the top decile, hold stocks with high momentum and low volatility, exclude stocks that fail a screen. Many such rules are Boolean functions of threshold statements about signal ranks, and writing them this way shows both what a rule holds and how often it trades. A Walsh decomposition splits the return of any rule into exposures to individual signals and to interactions among signals; an intersection and a union of the same two screens, for example, have opposite interaction exposures. The same coefficients, together with the persistence of the underlying statements, determine turnover. Interactions are less persistent than their components, so rules that load on higher-order interactions trade more, and thresholds discard information, so at equal risk threshold implementations of interactions trade more than smooth ones. In U.S.\ stocks from 1973 to 2024, these predictions hold closely: predicted and realized persistence have correlations of at least 0.999, predicted turnover matches realized turnover with a slope of 1.02, and the same interactions are more persistent as products of ranks than as threshold rules. Trading costs rise with order, and because premia fall with order, 58\% of single-signal portfolios have absolute premia above their quoted-spread costs over 1995--2024, against 24\% of two-signal and 6\% of three-signal interaction portfolios. Out of sample, interaction portfolios raise the Sharpe ratio of single-signal tangency portfolios, and over 1995--2024 they reject a six-factor model that conventional sorts on the same signals do not reject.

\medskip
\noindent\textbf{Keywords:} portfolio construction; decision rules; Boolean functions; interaction effects; turnover; transaction costs; factor investing
\end{abstract}

\newpage
\setcounter{page}{1}

\section{Introduction}\label{sec:intro}

To turn a cross-sectional signal into a portfolio, an investor needs a decision rule. A researcher ranks stocks on book-to-market and buys the top decile. A factor averages portfolios formed from intersections of median and 30/70 sorts. A quality screen keeps firms that pass at least eight of nine accounting tests \citep{Piotroski2000}, and an implemented portfolio first removes halted or illiquid stocks. The rule determines which exposures the portfolio carries and how often it trades, and reasonable rules built on the same signal can produce materially different returns \citep{HXZ2020,Menkveld2024,WWW2024}. Yet rules are usually treated as details of portfolio construction.

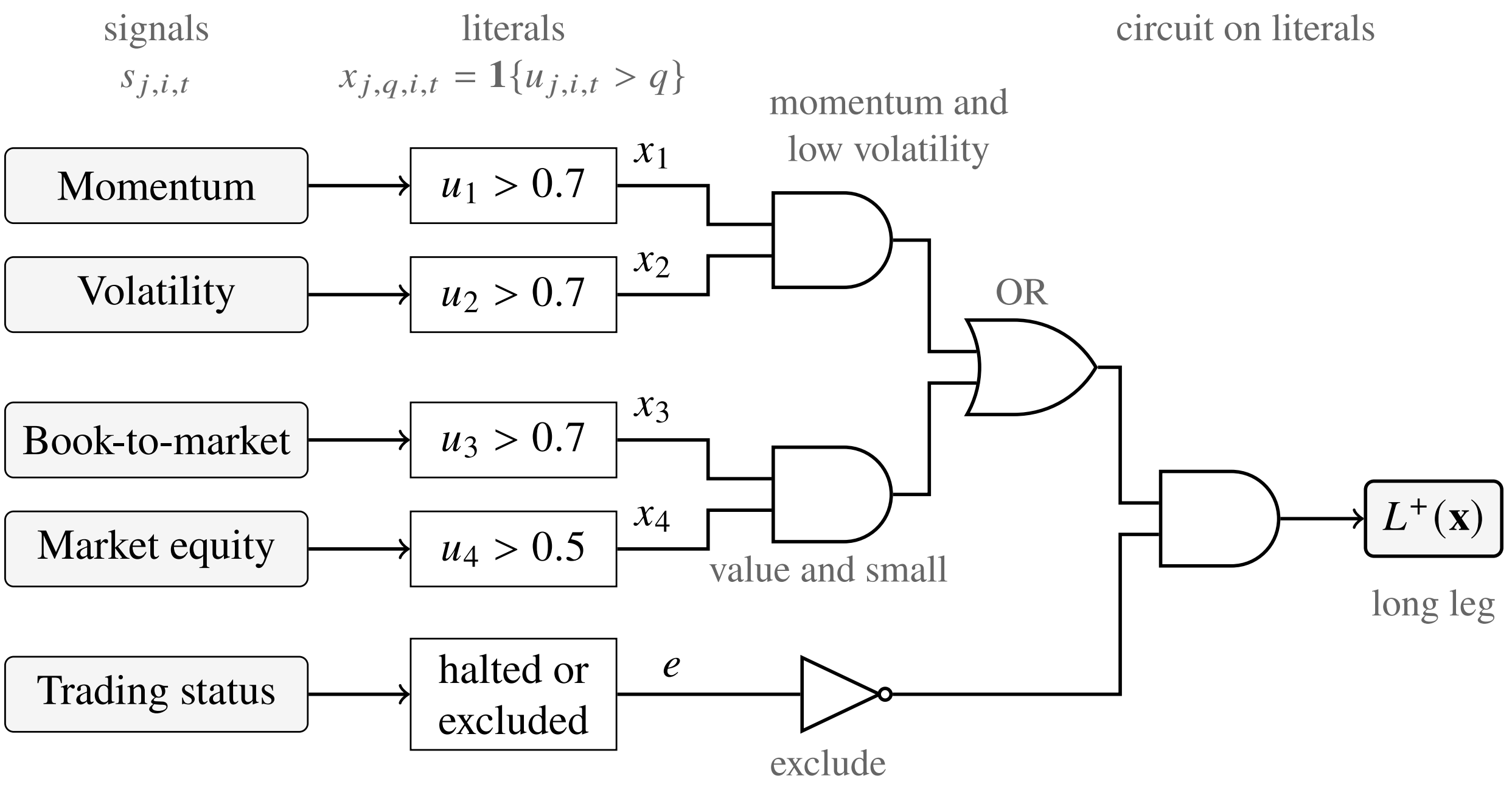
\begin{figure}[t]
\centering
\resizebox{\textwidth}{!}{\begin{tikzpicture}[
    circuit logic US,
    circuit symbol unit=5.5pt,
    every circuit symbol/.style={thick, fill=white},
    logic gate input sep=0.26cm,
    font=\small,
    sig/.style={draw, rounded corners=2pt, minimum width=25mm, minimum height=6.2mm, fill=black!4},
    cmp/.style={draw, minimum width=17mm, minimum height=6.2mm, fill=white, align=center},
    wire/.style={thick},
    note/.style={font=\footnotesize, text=black!60, align=center}
  ]
  \def\ya{3.0} \def\yb{2.1} \def\yc{0.9} \def\yd{0.0} \def\ye{-1.2}
  \foreach \y/\s/\c/\n in {\ya/Momentum/{$u_{1}> 0.7$}/1, \yb/Volatility/{$u_{2}> 0.7$}/2,
                           \yc/Book-to-market/{$u_{3}> 0.7$}/3, \yd/Market equity/{$u_{4}> 0.5$}/4,
                           \ye/Trading status/{halted or\\ excluded}/5} {
    \node[sig] (s\n) at (0,\y) {\s};
    \node[cmp] (c\n) at (2.95,\y) {\c};
    \draw[wire,->] (s\n.east) -- (c\n.west);
  }
  \node[and gate, inputs=nn] (a1) at (5.55, 2.55) {};
  \node[and gate, inputs=nn] (a2) at (5.55, 0.45) {};
  \node[or gate,  inputs=nn] (o1) at (7.15, 1.50) {};
  \node[not gate]            (n1) at (5.55,-1.20) {};
  \node[and gate, inputs=nn] (a3) at (8.75, 0.25) {};
  \draw[wire] (c1.east) -- node[above, pos=0.4] {$x_{1}$} ++(0.75,0) |- (a1.input 1);
  \draw[wire] (c2.east) -- node[above, pos=0.4] {$x_{2}$} ++(0.75,0) |- (a1.input 2);
  \draw[wire] (c3.east) -- node[above, pos=0.4] {$x_{3}$} ++(0.75,0) |- (a2.input 1);
  \draw[wire] (c4.east) -- node[above, pos=0.4] {$x_{4}$} ++(0.75,0) |- (a2.input 2);
  \draw[wire] (c5.east) -- node[above, pos=0.3] {$e$} (n1.input);
  \draw[wire] (a1.output) -- ++(0.3,0) |- (o1.input 1);
  \draw[wire] (a2.output) -- ++(0.3,0) |- (o1.input 2);
  \draw[wire] (o1.output) -| ($(a3.input 1)+(-0.32,0)$) -- (a3.input 1);
  \draw[wire] (n1.output) -| ($(a3.input 2)+(-0.32,0)$) -- (a3.input 2);
  \node[draw, thick, rounded corners=2pt, minimum height=6.2mm, fill=black!4] (out) at (10.55, 0.25) {$L^{+}(\mathbf{x})$};
  \draw[wire,->] (a3.output) -- (out.west);
  \node[note, anchor=south] at (6.05, 3.0) {momentum and\\ low volatility};
  \node[note, anchor=north] at (5.55, 0.08) {value and small};
  \node[note, anchor=south] at (7.15, 1.88) {OR};
  \node[note, anchor=north] at (5.55,-1.52) {exclude};
  \node[note, anchor=north] at (10.55,-0.20) {long leg};
  \node[note, anchor=north] at (0, 4.55)    {signals\\ $s_{j,i,t}$};
  \node[note, anchor=north] at (2.95, 4.55) {literals\\ $x_{j,q,i,t}=\mathbf{1}\{u_{j,i,t}>q\}$};
  \node[note, anchor=north] at (9.0, 4.55)  {circuit on literals};
\end{tikzpicture}}
\caption{A decision rule as a logic circuit. Each signal is ranked in the cross-section and compared with a threshold, producing a literal $x_{j,q,i,t}=\1\{u_{j,i,t}>q\}$; signals are oriented so that the favorable side has high rank (for volatility, low volatility). The long leg $L^{+}$ holds stocks with high momentum and low volatility, or with high book-to-market and small size, provided that the stock is neither halted nor on an exclusion list ($e=0$). A short leg $L^{-}$ is built the same way, and the portfolio weight of stock $i$ is proportional to $\omega_i\,[L^{+}(\x_i)-L^{-}(\x_i)]$, where $\omega_i$ is a base weight such as capped market value.}
\label{fig:circuit}
\end{figure}

This paper studies the rules themselves. Many common rules are Boolean functions of literals, which are statements such as ``the cross-sectional rank of signal $j$ exceeds $q$.'' Sorts, double sorts, screens, exclusions, and the leaves of decision trees all take this form, and \cref{fig:circuit} draws one such rule as a logic circuit. Writing a rule this way is exact and imposes no assumptions on returns. The representation is useful because one set of coefficients describes both the exposures of a rule and its turnover.

These coefficients come from the Walsh expansion, the Fourier expansion of Boolean functions \citep{ODonnell2014}. The expansion splits the return of any rule, month by month, into exposures to individual signals and exposures to interactions among signals. An intersection of two screens and their union, for example, have the same exposure to each signal but opposite exposures to the interaction. The intersection bets that the two signals reinforce each other; the union bets that either one is enough. Besides their standard $2\times3$ factors, \citet{FF2015} build factors from a $2\times2\times2\times2$ sort on size, value, profitability, and investment. In that sort, the four factors are the first-order (single-signal) coordinates, and the eleven interaction coordinates lie outside their span. Because the return of every rule built on the literals of $k$ signals at given thresholds combines portfolios that involve at most $k$ signals, testing these portfolios tests all such rules at once.

The same coefficients determine turnover. With equal base weights, a rule trades only when its value for a stock changes, which requires a change in the stock's literals. If literals are persistent and move independently across signals, the persistence of an interaction is the product of the persistences of its components. Two literals with a monthly persistence of 0.9 produce an interaction with persistence 0.81, and three produce 0.73. The one-way turnover of a leg equals the share of the universe it does not hold times a weighted average of one minus these persistences, with weights given by the decomposition. Rules that load on higher-order interactions therefore trade more.

The paper makes three contributions. The main one is the framework. It applies standard tools from the analysis of Boolean functions, including the Walsh expansion, Parseval's identity, noise sensitivity, and Sheppard's arcsine formula \citep{ODonnell2014}, to portfolio rules, where they connect the logic of a rule to its exposures, its turnover, and its trading costs. The return of every rule is an exact sum of first-order and interaction portfolios (\cref{prop:spectral}), so tests on these portfolios cover every rule built on the same literals (\cref{cor:span}). The turnover of a rule follows from the same coefficients and the persistence of its literals (\cref{cor:profile}), with closed forms for AND, OR, and threshold rules (\cref{prop:gates} and Online Appendix~\ref{oa:proofs}). Cutting a persistent signal at a threshold lowers its persistence, and the loss compounds in interactions (\cref{prop:discrete}), so at equal risk smooth implementations of interactions trade less than thresholds.

The second contribution is to test these predictions, and the data confirm them closely. In U.S.\ stocks from 1973 to 2024 with twelve well-known signals, predicted and realized persistence have a correlation of at least 0.999 within each order, and predicted turnover matches the realized turnover of 286 AND, OR, and majority rules with a slope of 1.02. The same exposures are more persistent when built from smooth ranks than from thresholds, as \cref{prop:discrete} predicts: second-order interactions have a mean persistence of 0.63 with median literals, 0.69 with three thresholds, and 0.73 with ranks, against predicted values of 0.63, 0.68, and 0.72, and the difference between ranks and median literals widens from 0.10 at second order to 0.13 at third order. At equal risk, rank-product interactions therefore trade less than median interactions under both quadratic and proportional trading costs. Because turnover rises with order, so do trading costs: over 1995--2024, the mean quoted-spread cost per unit of gross exposure rises from 4.8 basis points per month at first order to 7.5 and 9.5 at second and third order. Premia fall with order at the same time, so 58\% of first-order portfolios have absolute premia above their costs, against 24\% of second-order and 6\% of third-order portfolios.

The third contribution is to show what the decomposition reveals about expected returns. Interaction portfolios carry information beyond first-order portfolios built from the same signals: block GRS tests reject zero intercepts of the second-order portfolios on the first-order ones ($p<0.001$), and adding second-order portfolios raises the Sharpe ratio of an out-of-sample tangency portfolio from 0.18 to 0.48 over 2005--2024, and from 0.11 to 0.32 after trading costs. Because the portfolios of order at most $k$ span every rule on $k$ signals, a test on them tests all such rules at once, and in recent samples they are also sharper test assets than conventional sorts. In samples that start in 1995, 2000, or 2005, conventional sorts on our signals do not reject the five-factor model of \citet{FF2015} augmented with the momentum factor of \citet{Carhart1997} (FF6), whereas the first- and second-order portfolios of the decomposition do; since 1995, they also reject the $q^5$ model of \citet{HMXZ2021}, which conventional sorts do not. For portfolio managers, these results give a simple way to evaluate the interaction exposures of a rule before trading it (\cref{sec:design}).

Machine-learning methods for the cross-section \citep{GKX2020,BPZ2025,CFHH2025} search for the rule that predicts best, whereas we take a rule as given and ask what it holds and what it costs to trade. Online Appendix~\ref{oa:search} uses an exhaustive search over sixteen million rules to show why the two questions differ.

The paper relates to several literatures. The Fourier analysis of Boolean functions is a mature field of theoretical computer science and probability \citep{BKS1999,ODonnell2014}, and the closely related Hoeffding decomposition is a standard tool of statistics \citep{Hoeffding1948,EfronStein1981}. \citet{CCFS2020} show that characteristic-sorted portfolios are nonparametric partitioning estimators; the Boolean representation adds an orthogonal coordinate system on the partition cells and a notion of order. Tree methods \citep{GKX2020,BPZ2025,CFHH2025} learn conjunctions of threshold conditions, and their leaves are one class of rules in our framework. Evidence that interactions matter \citep{Asness1997,HLS2000,Nagel2005,SYY2015,FNW2020,MullerSchmickler2025} and that shrinkage is needed when the set of candidate assets is large \citep{KNS2020,KMZ2024} motivates the exposures we study and the ridge estimates in our out-of-sample tests. On implementation, we build on work on trading costs and their mitigation \citep{KorajczykSadka2004,NovyMarxVelikov2016,DNV2023,ChenVelikov2023}, on netting across characteristics \citep{DMNU2020}, and on dynamic trading with costs \citep{GarleanuPedersen2013}. We add a direct link between the trading frequency of a rule, its interaction structure, and the persistence of its signals, and our treatment of searched rules (Online Appendix~\ref{oa:search}) draws on the data-snooping literature \citep{White2000,RomanoWolf2005,HLZ2016,CGS2020}.

\Cref{sec:rules,sec:walsh,sec:turnover} develop the framework, \cref{sec:data,sec:costs} test its predictions, \cref{sec:economics} shows what it reveals about expected returns, and \cref{sec:design} draws the implications for portfolio construction.

\section{Decision rules as Boolean functions}\label{sec:rules}

\subsection{Literals and rules}

At the end of month $t$ the universe $\U_t$ contains $N_t$ stocks. Signal $j\in\{1,\dots,J\}$ takes value $s_{j,i,t}$ for stock $i$ and is oriented so that higher values are associated with higher average returns. Its cross-sectional rank $u_{j,i,t}\in(0,1]$ is the rank of $s_{j,i,t}$ divided by $N_t$. A literal is an indicator
\begin{equation}
x_{j,q,i,t}=\1\{u_{j,i,t}>q\},\qquad q\in(0,1).
\end{equation}
It answers a yes-or-no question about one signal: is the stock above the $q$-quantile of signal $j$ this month? With $q=0.9$ it marks the top decile, and with $q=1/2$ the favorable half. Its complement $\neg x_{j,q,i,t}=1-x_{j,q,i,t}$ marks the stocks at or below the quantile. Stacking the $n$ literals in use gives the vector $\x_{i,t}\in\{0,1\}^n$. A decision rule is a function $f:\{0,1\}^n\to\mathbb{R}$ of this vector. A rule with values in $\{0,1\}$ defines one leg of a portfolio, and a rule with values in $\{-1,0,1\}$ defines a long--short portfolio $L^{+}-L^{-}$.

Three families of rules appear throughout. For a set $G$ of $k$ literals, the AND rule holds a stock when every literal in $G$ equals one, the OR rule holds it when at least one does, and the threshold rule $\1\{\sum_{j\in G}x_j\ge c\}$ holds it when at least $c$ of them do. AND and OR are the threshold rules with $c=k$ and $c=1$, and the majority rule has $c=(k+1)/2$ for odd $k$.

\Cref{tab:circuits} writes familiar constructions in this notation. A dependent sort is a two-level decision tree whose second split uses ranks within the first group. The leaves of the asset-pricing trees of \citet{BPZ2025} and the panel trees of \citet{CFHH2025} are conjunctions of threshold conditions, and a set of leaves is a disjunction of such conjunctions. The anomaly counts of \citet{EMP2018} are sums of literals, so sorting on them gives threshold rules such as the F-score rule. Mixing two single-signal portfolios is itself a rule: it gives weight two to the stocks in both portfolios and weight one to the stocks in exactly one.

\begin{table}[t]\centering
\caption{Familiar portfolio rules as Boolean functions}\label{tab:circuits}
\begin{threeparttable}\small
\begin{tabular}{ll}\toprule
Rule & Boolean form \\ \midrule
Top-decile portfolio of signal $j$ & $x_{j,0.9}$ \\
Top-minus-bottom decile spread & $x_{j,0.9}-\neg x_{j,0.1}$ \\
Cell of an independent $2\times2$ sort & $x_{1,1/2}\wedge x_{2,1/2}$ \\
High momentum and low volatility & $x_{\mathrm{MOM},q}\wedge x_{\mathrm{VOL},q}$ \\
Tree leaf along one path & $x_{1,q_1}\wedge\neg x_{2,q_2}\wedge x_{3,q_3}$ \\
Set of tree leaves & disjunction of leaves \\
Exclusion of halted stocks & $L\wedge\neg e$ \\
F-score of at least eight & $\1\{\sum_{j=1}^{9}x_j\ge8\}$ \\
Equal mix of two single-signal portfolios & $x_1+x_2=(x_1\vee x_2)+(x_1\wedge x_2)$ \\
\bottomrule\end{tabular}
\begin{tablenotes}\footnotesize
\item Subscripts $i$ and $t$ are omitted. $\wedge$, $\vee$, and $\neg$ denote AND, OR, and NOT; for zero--one variables $x\wedge y=xy$, $x\vee y=x+y-xy$, and $\neg x=1-x$. The volatility signal is oriented so that its literal marks low volatility, and $e=1$ flags an excluded stock.
\end{tablenotes}\end{threeparttable}\end{table}

\subsection{From rules to portfolios}

Let $\omega_{i,t}\ge0$ with $\sum_{i\in\U_t}\omega_{i,t}=1$ be base weights (equal, value, or capped value), $r_{i,t+1}$ the excess return from $t$ to $t+1$, $\bar r_{t+1}=\sum_i\omega_{i,t}r_{i,t+1}$ the base-weighted market return, and $\tilde r_{i,t+1}=r_{i,t+1}-\bar r_{t+1}$. The managed portfolio of rule $f$ is
\begin{equation}\label{eq:managed}
F^f_{t+1}=\sum_{i\in\U_t}\omega_{i,t}\,f(\x_{i,t})\,\tilde r_{i,t+1}=\sum_{i\in\U_t}\omega_{i,t}\bigl(f(\x_{i,t})-\bar f_t\bigr)r_{i,t+1},
\end{equation}
where $\bar f_t=\sum_i\omega_{i,t}f(\x_{i,t})$ is the base-weighted average of the rule. The managed portfolio is a zero-investment portfolio that is long the stocks on which the rule exceeds its average and short the others, in proportion to their base weights. The analysis below uses only two of its properties: adding a constant to $f$ leaves $F^f$ unchanged, and $F^f$ is linear in the rule, $F^{af+bg}=aF^f+bF^g$.

The managed portfolio differs from the conventional portfolio only by a scale known at $t$. Let $L$ be a leg, $\beta_t=\sum_i\omega_{i,t}L(\x_{i,t})$ its coverage (the share of base weight in the stocks it holds), and $R_{t+1}(L)=\beta_t^{-1}\sum_{i:L(\x_{i,t})=1}\omega_{i,t}r_{i,t+1}$ its conventional base-weighted return. Then
\begin{equation}\label{eq:leg}
F^L_{t+1}=\beta_t\bigl(R_{t+1}(L)-\bar r_{t+1}\bigr).
\end{equation}
The managed return of a leg is its return in excess of the base-weighted market return, multiplied by its coverage. For a long--short portfolio whose two legs have the same coverage, \eqref{eq:leg} gives the conventional spread times $\beta_t$, so the two versions have the same $t$-statistic when $\beta_t$ is constant.

\section{The interaction decomposition}\label{sec:walsh}

The tools in this section are standard in the Fourier analysis of Boolean functions \citep[Chapters 1 and 8]{ODonnell2014}. We restate them in portfolio terms, where the coefficients of the expansion become the exposures of a rule.

\subsection{Walsh characters}

Consider first one literal per signal at the median, so that $n=J$ and each literal equals one for half of the universe. Recode each literal as a sign, $2x_j-1$, which is $+1$ on the favorable half and $-1$ otherwise. For a set of literals $S\subseteq\{1,\dots,n\}$, the Walsh character is the product of their signs,
\[
\chi_S(\x)=\prod_{j\in S}(2x_j-1),\qquad \chi_\emptyset\equiv1 ,
\]
and we call $|S|$ its order. A character of order one is the sign of one signal. A character of order two equals $+1$ when the stock is on the same side of both signals and $-1$ when the two sides disagree.

The characters form an orthonormal basis. To see this, equip $\{0,1\}^n$ with the uniform distribution, under which the literals are independent fair coins. Because $(2x_j-1)^2=1$, the product of two characters is again a character, $\chi_S\chi_{S'}=\chi_{S\triangle S'}$, where $S\triangle S'$ contains the literals that belong to exactly one of the two sets. Every character other than $\chi_\emptyset$ has mean zero, so $2^{-n}\sum_{\mathbf z}\chi_S(\mathbf z)\chi_{S'}(\mathbf z)=\1\{S=S'\}$. The $2^n$ characters therefore form a basis of the $2^n$-dimensional space of functions on $\{0,1\}^n$, and projecting a rule on them gives the Walsh expansion
\begin{equation}\label{eq:walsh}
f(\x)=\sum_{S}\hat f(S)\,\chi_S(\x),\qquad \hat f(S)=2^{-n}\sum_{\mathbf z\in\{0,1\}^n}f(\mathbf z)\chi_S(\mathbf z).
\end{equation}
The expansion holds exactly at every point; the uniform distribution enters only the formula for the coefficients. The coefficient $\hat f(S)$ is the exposure of the rule to the interaction among the literals in $S$, and the order of the rule is the largest $|S|$ with $\hat f(S)\neq0$.

\subsection{AND versus OR}

\Cref{tab:walsh2} works out the case of two literals. Writing $x_j=\tfrac12(1+\chi_{\{j\}})$ and multiplying out gives
\begin{align*}
x_1\wedge x_2&=x_1x_2=\tfrac14\bigl(1+\chi_{\{1\}}+\chi_{\{2\}}+\chi_{\{1,2\}}\bigr),\\
x_1\vee x_2&=x_1+x_2-x_1x_2=\tfrac14\bigl(3+\chi_{\{1\}}+\chi_{\{2\}}-\chi_{\{1,2\}}\bigr).
\end{align*}
The intersection and the union have the same first-order exposures and differ only in the sign of the interaction exposure. The exclusive or, which holds a stock that is favorable on exactly one signal, has no first-order exposure and is a pure interaction, $\tfrac12(1-\chi_{\{1,2\}})$.

\begin{table}[t]\centering
\caption{Interaction decomposition of rules on two median literals}\label{tab:walsh2}
\begin{threeparttable}\small
\begin{tabular}{cccccccc}\toprule
\multicolumn{8}{l}{\emph{Panel A. Values}}\\
$x_1$ & $x_2$ & $\chi_{\{1\}}$ & $\chi_{\{2\}}$ & $\chi_{\{1,2\}}$ & $x_1\wedge x_2$ & $x_1\vee x_2$ & $x_1\oplus x_2$ \\ \midrule
0 & 0 & $-1$ & $-1$ & $+1$ & 0 & 0 & 0 \\
0 & 1 & $-1$ & $+1$ & $-1$ & 0 & 1 & 1 \\
1 & 0 & $+1$ & $-1$ & $-1$ & 0 & 1 & 1 \\
1 & 1 & $+1$ & $+1$ & $+1$ & 1 & 1 & 0 \\
\bottomrule\end{tabular}

\medskip
\begin{tabular}{lccccc}\toprule
\multicolumn{6}{l}{\emph{Panel B. Exposures (Walsh coefficients) and interaction share}}\\
Rule & $\hat f(\emptyset)$ & $\hat f(\{1\})$ & $\hat f(\{2\})$ & $\hat f(\{1,2\})$ & Interaction share \\ \midrule
AND, $x_1\wedge x_2$ & $1/4$ & $1/4$ & $1/4$ & $1/4$ & $1/3$ \\
OR, $x_1\vee x_2$ & $3/4$ & $1/4$ & $1/4$ & $-1/4$ & $1/3$ \\
Exclusive or, $x_1\oplus x_2$ & $1/2$ & 0 & 0 & $-1/2$ & 1 \\
\bottomrule\end{tabular}
\begin{tablenotes}\footnotesize
\item Each coefficient is $\hat f(S)=\tfrac14\sum_{\mathbf z}f(\mathbf z)\chi_S(\mathbf z)$, the average over the four rows of Panel A of the rule times the character. For example, $\hat f(\{1,2\})$ for OR is $\tfrac14(0-1-1+1)=-\tfrac14$. The interaction share is defined in \cref{def:profile}.
\end{tablenotes}\end{threeparttable}\end{table}

The same algebra applies to an intersection of any set $G$ of $k$ literals:
\begin{equation}\label{eq:andk}
\prod_{j\in G}x_j=2^{-k}\prod_{j\in G}\bigl(1+\chi_{\{j\}}\bigr)=2^{-k}\sum_{S\subseteq G}\chi_S .
\end{equation}
An intersection spreads its exposure equally over the $2^k$ characters built from its literals, $\binom kd$ of which have order $d$. A union of the same literals has exposures of the same size, because $\vee_{j\in G}x_j=1-\prod_{j\in G}(1-x_j)$ and $1-x_j=\tfrac12(1-\chi_{\{j\}})$; its exposures have the same sign as those of the intersection at odd orders and the opposite sign at even orders. To summarize where the exposure of a rule sits, we use its interaction profile.

\begin{definition}[Interaction profile]\label{def:profile}
For a rule $f$ that is not constant, the interaction profile is $w_S(f)=\hat f(S)^2/\sum_{T\neq\emptyset}\hat f(T)^2$ for $S\neq\emptyset$. The interaction share of $f$ is $\sum_{|S|\ge2}w_S(f)$, the part of the variation of the rule under independent literals that comes from interactions.
\end{definition}

The profile is a probability distribution over sets of literals; by Parseval's identity, $\sum_{T\neq\emptyset}\hat f(T)^2$ is the variance of $f(\x)$ under independent fair literals. The intersection and the union of two literals both have interaction share $1/3$, and the exclusive or has share one. By \eqref{eq:andk}, an intersection or union of $k$ literals has interaction share $1-k/(2^k-1)$, which is 0.57 for three literals and 0.84 for five. A majority rule on three literals has share $1/4$. Together with the persistence of the literals, the profile determines how much a rule trades (\cref{sec:turnover}).

\subsection{Decomposing rule returns}

The managed portfolio of a character, $\Pi^S_{t+1}=F^{\chi_S}_{t+1}$, is the character portfolio of $S$. We call it a first-order portfolio when $|S|=1$ and an interaction portfolio when $|S|\ge2$. By \eqref{eq:leg}, a first-order portfolio is the high-minus-low spread of one signal, scaled by $2\beta_t(1-\beta_t)$, which equals one half when $\beta_t=\tfrac12$. A second-order portfolio is a difference-in-differences. When the four cells of a $2\times2$ sort on signals $j$ and $k$ have equal weight,
\[
\Pi^{\{j,k\}}_{t+1}=\tfrac14\Bigl[\bigl(R^{HH}_{t+1}-R^{HL}_{t+1}\bigr)-\bigl(R^{LH}_{t+1}-R^{LL}_{t+1}\bigr)\Bigr],
\]
where $R^{c}_{t+1}$ is the return of cell $c$ and the first letter refers to signal $j$. The portfolio measures how much more the spread on signal $k$ pays among stocks that are favorable on signal $j$ than among stocks that are unfavorable on it. When a literal does not split the base weight in half, as with capped value weights, an interaction portfolio also carries a scaled copy of lower-order spreads. Equal weights remove this component from second-order portfolios, and regressing higher-order portfolios on lower-order ones removes it at every order (Online Appendix~\ref{oa:proofs}).

\begin{proposition}[Return decomposition]\label{prop:spectral}
For every $f:\{0,1\}^n\to\mathbb{R}$ and every $t$,
\[
F^f_{t+1}=\sum_{S\neq\emptyset}\hat f(S)\,\Pi^S_{t+1}.
\]
\end{proposition}

\begin{proof}
Evaluate \eqref{eq:walsh} at the literal vector of each stock, multiply by $\omega_{i,t}\tilde r_{i,t+1}$, and sum over stocks. Exchanging the two finite sums,
\[
F^f_{t+1}=\sum_i\omega_{i,t}\sum_S\hat f(S)\chi_S(\x_{i,t})\tilde r_{i,t+1}=\sum_S\hat f(S)\sum_i\omega_{i,t}\chi_S(\x_{i,t})\tilde r_{i,t+1}=\sum_S\hat f(S)\,\Pi^S_{t+1}.
\]
The term with $S=\emptyset$ vanishes because $\chi_\emptyset\equiv1$ and $\sum_i\omega_{i,t}\tilde r_{i,t+1}=0$.
\end{proof}

The coefficients depend only on the rule, not on the data, so the decomposition holds month by month and for any joint distribution of signals and returns. Correlated signals make the character portfolios correlated with one another but do not change the decomposition. For the rules of \cref{tab:walsh2} it reads
\[
F^{x_1\wedge x_2}_{t+1}=\tfrac14\bigl(\Pi^{\{1\}}_{t+1}+\Pi^{\{2\}}_{t+1}+\Pi^{\{1,2\}}_{t+1}\bigr),\qquad
F^{x_1\vee x_2}_{t+1}=\tfrac14\bigl(\Pi^{\{1\}}_{t+1}+\Pi^{\{2\}}_{t+1}-\Pi^{\{1,2\}}_{t+1}\bigr),
\]
so an intersection and a union of the same two literals differ by a pure interaction bet, $\tfrac12\Pi^{\{1,2\}}_{t+1}$. The decomposition has three consequences for testing.

\begin{corollary}[Sorts, spans, and alphas]\label{cor:span}
Let $\F_k$ be the set of functions of $\x$ that depend on at most $k$ literals.
\begin{enumerate}[label=(\roman*),nosep]
\item $\spn\{F^f:f\in\F_k\}=\spn\{\Pi^S:1\le|S|\le k\}$.
\item If the character portfolios satisfy $\Pi^S_{t}=\alpha_S+\bm\delta_S'\mathbf G_t+\varepsilon_{S,t}$ for factors $\mathbf G_t$, then $F^f$ has intercept $\alpha_f=\sum_{S\neq\emptyset}\hat f(S)\alpha_S$.
\item Let $\bm\alpha$ and $\Sigma_\varepsilon$ collect the intercepts and residual covariance of the character portfolios with $1\le|S|\le k$. If $\Sigma_\varepsilon$ is nonsingular, every portfolio in the span in (i) satisfies $\alpha_P^2/\sigma^2(\varepsilon_P)\le\bm\alpha'\Sigma_\varepsilon^{-1}\bm\alpha$, with equality for weights proportional to $\Sigma_\varepsilon^{-1}\bm\alpha$.
\end{enumerate}
\end{corollary}

Part (i) says that a rule cannot load on a literal it ignores: if $f$ does not depend on literal $j$, flipping $x_j$ leaves $f$ unchanged but reverses the sign of every character that contains $j$. Part (ii) follows from the decomposition, and part (iii) from the Cauchy--Schwarz inequality (Online Appendix~\ref{oa:proofs}). The ratio in part (iii) is the squared Sharpe ratio of a portfolio whose factor exposures have been hedged away, and the bound is the largest value any combination of character portfolios can reach. The test of \citet{GRS1989} (GRS) is a scaled sample version of the same quadratic form. By part (i), the character portfolios of order at most $k$ span the managed portfolio of every rule that sorts, screens, or combines $k$ of the signals at the chosen breakpoints, so a GRS test on these portfolios tests all such rules at once. For rules on a fixed set of literals, this addresses the choice of test assets discussed by \citet{DanielTitman1997} and \citet{LNS2010}.

\begin{example}[The $2\times2\times2\times2$ sort]\label{ex:ff}
Let $g(\mathbf z)$ be the capped value-weighted return of the cell $\mathbf z\in\{0,1\}^4$ defined by median literals on size, book-to-market, profitability, and investment, and let $\hat g(A)=2^{-4}\sum_{\mathbf z}g(\mathbf z)\chi_A(\mathbf z)$. In the $2\times2\times2\times2$ construction that \citet{FF2015} report alongside their standard $2\times3$ factors, the factor for signal $h$ is the average of the eight cells with $z_h=1$ minus the average of the eight cells with $z_h=0$, which equals $2\hat g(\{h\})$. The interaction spread of two signals $h\neq h'$, half the difference between the $h$ spread among cells with $z_{h'}=1$ and that among cells with $z_{h'}=0$, equals $2\hat g(\{h,h'\})$. The four factors are therefore the first-order coordinates of the sort, and the eleven coefficients with $|A|\ge2$ are interaction coordinates outside the span of the four factors. The coefficients weight the sixteen cells equally; they equal the character portfolios $\Pi^A$ when every cell has the same total base weight (Online Appendix~\ref{oa:proofs}).
\end{example}

\subsection{Several thresholds and smooth scores}

We call the character portfolios built from median literals median portfolios. The same construction works when a signal is cut at several thresholds. Its literals then place each stock in one of $m_j+1$ ordered buckets $\kappa_j$ with nominal probabilities $\pi_j(\kappa)$. The signs are no longer orthogonal, so we orthogonalize the literals by Gram--Schmidt under $\pi_j$. This gives functions $\psi_{j,0}\equiv1,\psi_{j,1},\dots,\psi_{j,m_j}$ of $\kappa_j$, which we multiply across signals, $\psi_{\bm a}(\bm\kappa)=\prod_j\psi_{j,a_j}(\kappa_j)$. We use thresholds at the 20th, 50th, and 80th percentiles. The three functions of a signal are then the median sign, a comparison of the top quintile with the rest of the upper half, and a comparison of the 20--50 bucket with the bottom quintile (Online Appendix~\ref{oa:proofs}). The products play the role of characters, and we call their managed portfolios three-threshold portfolios. The order of $\psi_{\bm a}$ is the number of signals with $a_j\neq0$, so twelve signals give 36 first-order and 594 second-order three-threshold portfolios. \Cref{prop:spectral,cor:span} hold with these products in place of characters, and grouping terms by the signals involved gives the Hoeffding or Efron--Stein decomposition \citep{Hoeffding1948,EfronStein1981}.

A rule can also use smooth scores instead of thresholds. Rank products $(u_j-\tfrac12)(u_k-\tfrac12)$, the interaction terms used by \citet{KNS2020}, are limits of combinations of second-order products $\psi_{\bm a}$ as the thresholds become dense. Replacing each rank by the mean rank of its bucket turns a rank product into such a combination, with an error no larger than the width of the widest bucket. Median portfolios, three-threshold portfolios, and rank products therefore target the same second-order exposures with increasingly fine resolution.

\section{Interactions and turnover}\label{sec:turnover}

With equal base weights, a rule trades only when its value for a stock changes. How often this happens depends on the interaction profile of the rule and on the persistence of its literals. The probability that a Boolean function changes value when its inputs are partly redrawn is its noise sensitivity, and the formula for noise sensitivity in terms of the Walsh coefficients is a standard result \citep[][Chapters 2 and 8]{BKS1999,ODonnell2014}. We use it to derive the turnover of portfolio rules.

\subsection{Turnover and the movement of literals}

\begin{lemma}[Turnover identity]\label{lem:turnover}
Let $f$ be Boolean, let the base weights be equal, and let the universe contain the same $N$ stocks at $t$ and $t+1$. Then the two-sided turnover of the portfolio with target weights $f(\x_{i,t})/N$, measured as the sum of the absolute changes in target weights, equals the fraction of stocks whose value of $f$ changes, $N^{-1}\sum_i\1\{f(\x_{i,t})\neq f(\x_{i,t+1})\}$.
\end{lemma}

\begin{proof}
The target weight of stock $i$ changes by $\bigl(f(\x_{i,t+1})-f(\x_{i,t})\bigr)/N$, and $|a-b|=\1\{a\neq b\}$ for $a,b\in\{0,1\}$.
\end{proof}

Turnover therefore counts entries and exits. With value weights it also reflects changes in the base weights, which the cost calculations in \cref{sec:costs} include. To relate turnover to the structure of a rule, we need a model of how literals move.

\begin{assumption}[Persistent literals]\label{ass:markov}
Use one literal per signal. For each stock the pairs $(x_{j,t},x_{j,t+1})$ are independent across signals $j$. Each $x_{j,t}$ and $x_{j,t+1}$ is Bernoulli$(p_j)$, and $\mathrm{corr}(x_{j,t},x_{j,t+1})=\rho_j$, with $\rho_j\ge-\min\{p_j/(1-p_j),(1-p_j)/p_j\}$.
\end{assumption}

A pair of Bernoulli$(p_j)$ variables with correlation $\rho_j$ has $\Pr[x_{j,t}=1,x_{j,t+1}=1]=p_j^2+\rho_jp_j(1-p_j)$, and the marginals fix the other three cells of the joint distribution; the lower bound on $\rho_j$ keeps all four nonnegative. The literal changes value with probability $2p_j(1-p_j)(1-\rho_j)$. A literal with $\rho_j=1$ never changes, one with $\rho_j=0$ is redrawn independently each month, and one with $\rho_j<0$, such as the short-term reversal literal, tends to change. For $\rho_j\ge0$ the model is equivalent to keeping the value of the literal with probability $\rho_j$ and otherwise redrawing it from Bernoulli$(p_j)$. Independence across signals is a simplification; \cref{sec:turnover-evidence} shows where it fails and that the model still predicts turnover closely.

When $p_j\neq1/2$ the sign $2x_j-1$ no longer has mean zero, so we standardize the literal, $\phi_j(x)=(x-p_j)/\sqrt{p_j(1-p_j)}$, and form the products $\phi_S=\prod_{j\in S}\phi_j$. Because the literals are independent across signals, $\E[\phi_S\phi_{S'}]$ factors into one term per literal, equal to $\E[\phi_j^2]=1$ for $j$ in both sets and to $\E[\phi_j]=0$ for $j$ in only one, so $\E[\phi_S\phi_{S'}]=\1\{S=S'\}$. The coefficients $\hat f_p(S)=\E[f(\x_t)\phi_S(\x_t)]$ generalize the Walsh coefficients and coincide with them for median literals.

\subsection{The persistence of interactions}

\begin{proposition}[Persistence of interactions]\label{prop:noise}
Under \cref{ass:markov}, with $\rho_S=\prod_{j\in S}\rho_j$:
\begin{enumerate}[label=(\roman*),nosep]
\item $\E[\phi_S(\x_t)\phi_{S'}(\x_{t+1})]=\1\{S=S'\}\,\rho_S$;
\item $\E[f(\x_t)f(\x_{t+1})]=\sum_S\rho_S\hat f_p(S)^2$ for every $f$;
\item if $f$ is Boolean, $\Pr[f(\x_t)\neq f(\x_{t+1})]=2\sum_{S\neq\emptyset}(1-\rho_S)\hat f_p(S)^2$.
\end{enumerate}
For median literals, the character $\chi_S$ changes sign with probability $(1-\rho_S)/2$.
\end{proposition}

\begin{proof}
(i) Because the pairs $(x_{j,t},x_{j,t+1})$ are independent across $j$, the expectation factors into one term per literal:
\[
\E[\phi_S(\x_t)\phi_{S'}(\x_{t+1})]=\prod_{j\in S\cap S'}\E[\phi_j(x_{j,t})\phi_j(x_{j,t+1})]\prod_{j\in S\setminus S'}\E[\phi_j(x_{j,t})]\prod_{j\in S'\setminus S}\E[\phi_j(x_{j,t+1})].
\]
Every factor in the last two products is zero, so the expectation vanishes unless $S=S'$. For $j\in S=S'$, $\phi_j$ standardizes a Bernoulli$(p_j)$ variable, so $\E[\phi_j(x_{j,t})\phi_j(x_{j,t+1})]=\mathrm{corr}(x_{j,t},x_{j,t+1})=\rho_j$.
(ii) Expand $f(\x_t)$ and $f(\x_{t+1})$ in the basis, multiply, and take expectations. By (i), only the terms with $S=S'$ survive, each with weight $\rho_S$.
(iii) For Boolean $f$, $\1\{f(\x_t)\neq f(\x_{t+1})\}=\bigl(f(\x_t)-f(\x_{t+1})\bigr)^2$, so
\[
\Pr[f(\x_t)\neq f(\x_{t+1})]=\E[f(\x_t)^2]+\E[f(\x_{t+1})^2]-2\E[f(\x_t)f(\x_{t+1})].
\]
Because $\x_{t+1}$ has the same distribution as $\x_t$, orthonormality gives $\E[f(\x_t)^2]=\E[f(\x_{t+1})^2]=\sum_S\hat f_p(S)^2$. Combining with (ii) yields $2\sum_S(1-\rho_S)\hat f_p(S)^2$, and the term with $S=\emptyset$ vanishes because $\rho_\emptyset=1$. For median literals, $\chi_S\in\{-1,1\}$, so $\E[\chi_S(\x_t)\chi_S(\x_{t+1})]=1-2\Pr[\chi_S(\x_t)\neq\chi_S(\x_{t+1})]$, and (i) gives the last claim.
\end{proof}

Part (i) is the core of the mechanism. A median character is a product of signs, and it changes sign when an odd number of its signs flip. With independent literals the correlations multiply, so persistence decays geometrically as literals are added. Part (iii) is the noise sensitivity of the rule, generalized to a separate persistence for each literal: the contribution of each character to turnover is its squared coefficient times one minus its persistence. Because $\rho_S$ falls with order when literals are positively persistent, rules with larger interaction shares trade more. How much a leg trades also depends on how much of the universe it holds.

\begin{corollary}[Turnover of a leg]\label{cor:profile}
Under \cref{ass:markov}, let $f$ be Boolean with coverage $\beta=\Pr[f(\x_t)=1]\in(0,1)$, interaction profile $w_S=\hat f_p(S)^2/[\beta(1-\beta)]$, and one-way turnover $\tau(f)=\Pr[f(\x_{t+1})=0\mid f(\x_t)=1]$, the fraction of the leg sold after one month.
\begin{enumerate}[label=(\roman*),nosep]
\item $\tau(f)=(1-\beta)\,\bar\tau(f)$, where $\bar\tau(f)=\sum_{S\neq\emptyset}w_S(1-\rho_S)$.
\item For any $f$, Boolean or not, $\E\bigl[\bigl(f(\x_{t+1})-f(\x_t)\bigr)^2\bigr]=2\sum_{S\neq\emptyset}(1-\rho_S)\hat f_p(S)^2$.
\end{enumerate}
\end{corollary}

\begin{proof}
(i) Because $f(\x_t)$ and $f(\x_{t+1})$ have the same distribution, exits and entries are equally likely: $\Pr[f(\x_t)=1,f(\x_{t+1})=0]=\beta-\Pr[f(\x_t)=1,f(\x_{t+1})=1]=\Pr[f(\x_t)=0,f(\x_{t+1})=1]$. Hence $\Pr[f(\x_t)\neq f(\x_{t+1})]=2\beta\,\tau(f)$, and \cref{prop:noise}(iii) gives $\tau(f)=\beta^{-1}\sum_{S\neq\emptyset}(1-\rho_S)\hat f_p(S)^2$. By Parseval's identity, the variance of $f$ is $\beta(1-\beta)=\sum_{S\neq\emptyset}\hat f_p(S)^2$, so the weights $w_S$ sum to one and (i) follows.
(ii) Expand the square and use stationarity, orthonormality, and \cref{prop:noise}(ii), as in the proof of \cref{prop:noise}(iii).
\end{proof}

Part (i) splits the turnover of a leg into two factors. The first, $1-\beta$, is the share of the universe outside the leg. The second, $\bar\tau(f)$, is the average of $1-\rho_S$ over the exposures of the rule, weighted by its interaction profile; it depends only on the logic of the rule and the persistence of its literals. An intersection and a union of the same median literals have the same profile and therefore the same $\bar\tau$, but the narrow intersection sells a much larger share of its holdings than the broad union. Part (ii) matters for large portfolios. When trading costs are quadratic in the size of a trade, as in models of price impact \citep{GarleanuPedersen2013}, the expected cost of any rule, including a smooth score, is proportional to the same quadratic form $\sum_S(1-\rho_S)\hat f_p(S)^2$.

\subsection{AND, OR, and threshold rules}

For AND and OR rules, the turnover of a leg also has a direct closed form. Let $a_j=p_j+\rho_j(1-p_j)$ be the probability that a literal equal to one stays at one, and $b_j=(1-\rho_j)p_j$ the probability that a literal equal to zero switches to one; both follow from the joint distribution above.

\begin{proposition}[AND and OR rules]\label{prop:gates}
Under \cref{ass:markov}, for a set $G$ of $k$ literals:
\begin{enumerate}[label=(\roman*),nosep]
\item $\tau(\mathrm{AND}_G)=1-\prod_{j\in G}a_j$;
\item $\tau(\mathrm{OR}_G)=\prod_{j\in G}(1-p_j)\,\bigl[1-\prod_{j\in G}(1-b_j)\bigr]\big/\bigl[1-\prod_{j\in G}(1-p_j)\bigr]$;
\item $\tau(\mathrm{AND}_{G\cup\{l\}})\ge\tau(\mathrm{AND}_G)$ for every literal $l\notin G$; if $p_j=p\in(0,1)$ and $\rho_j=\rho$ for all $j$, then $\tau(\mathrm{OR}_G)$ is non-increasing in $k$ and bounded by $(1-\rho)(1-p)$, the turnover of a single literal.
\end{enumerate}
\end{proposition}

Online Appendix~\ref{oa:proofs} proves the proposition and gives the turnover of threshold rules, including majority rules. An intersection keeps a stock only if every literal survives, so each added literal is one more way to leave. A union loses a stock only when all its literals are zero at the same time, which becomes rarer as the union grows when its literals share the same parameters. The result for unions does not hold in general. Adding a median literal with $\rho=0$ to one with $\rho=0.98$ raises the turnover of the leg from 0.01 to 0.17, because the new literal brings in stocks that leave again the next month.

\Cref{tab:gates} evaluates these results when every literal has $p=1/2$ and $\rho=0.9$. The interaction share and $\bar\tau$ rise with the number of literals and are the same for AND and OR rules, but coverage moves in opposite directions. An intersection of five literals holds 3\% of the universe and sells 23\% of its holdings every month; a union of five literals holds 97\% and sells less than 1\%. Majority rules hold half of the universe and have lower interaction shares and lower $\bar\tau$ than intersections or unions of the same size.

\begin{table}[t]\centering
\caption{Interaction profile and turnover of AND, OR, and majority rules when all literals have $p=1/2$ and $\rho=0.9$}\label{tab:gates}
\begin{threeparttable}\small
\begin{tabular}{ccccccccccc}\toprule
 & \multicolumn{2}{c}{AND and OR} & \multicolumn{2}{c}{AND} & \multicolumn{2}{c}{OR} & \multicolumn{3}{c}{Majority} \\ \cmidrule(lr){2-3}\cmidrule(lr){4-5}\cmidrule(lr){6-7}\cmidrule(lr){8-10}
$k$ & Int.\ share & $\bar\tau$ & Coverage & $\tau$ & Coverage & $\tau$ & Int.\ share & $\bar\tau$ & $\tau$ \\ \midrule
1 & 0.000 & 0.100 & 0.500 & 0.050 & 0.500 & 0.050 & 0.000 & 0.100 & 0.050 \\
2 & 0.333 & 0.130 & 0.250 & 0.098 & 0.750 & 0.033 & & & \\
3 & 0.571 & 0.163 & 0.125 & 0.143 & 0.875 & 0.020 & 0.250 & 0.143 & 0.071 \\
4 & 0.733 & 0.198 & 0.063 & 0.185 & 0.938 & 0.012 & & & \\
5 & 0.839 & 0.234 & 0.031 & 0.226 & 0.969 & 0.007 & 0.297 & 0.170 & 0.085 \\
\bottomrule\end{tabular}
\begin{tablenotes}\footnotesize
\item Int.\ share is the interaction share of \cref{def:profile}; $\bar\tau=\sum_Sw_S(1-\rho_S)$ is defined in \cref{cor:profile}; coverage is the fraction of the universe in the leg; and $\tau=(1-\text{coverage})\,\bar\tau$ is the fraction of the leg sold after one month. The majority rule is the threshold rule with $c=(k+1)/2$. All values follow from \cref{prop:gates}, \cref{cor:profile}, or the threshold-rule formula in Online Appendix~\ref{oa:proofs}, with $a=0.95$ and $b=0.05$.
\end{tablenotes}\end{threeparttable}\end{table}

\subsection{Reducing trading costs}\label{sec:levers}

\Cref{cor:profile} suggests three ways to reduce trading costs while keeping the main exposures of a rule. A buffer zone keeps a stock until its rank falls well below the entry threshold. It raises the persistence of each literal, and because persistence multiplies across literals, the gain is larger for interactions. With literal persistence $\rho$, an exposure of order $d$ has persistence $\rho^d$, so raising $\rho$ from 0.90 to 0.95 lowers $1-\rho^d$ from 0.10 to 0.05 at first order, from 0.19 to 0.10 at second order, and from 0.27 to 0.14 at third order.

Truncation drops high-order terms. Cutting the expansion at order $d$ gives $f^{\le d}=\sum_{|S|\le d}\hat f(S)\chi_S$, the closest function of order at most $d$ in mean square. By Parseval's identity the truncated rule keeps $\sum_{1\le|S|\le d}\hat f(S)^2$ of the variation of the rule, and by \cref{cor:profile}(ii) it removes $\sum_{|S|>d}(1-\rho_S)\hat f(S)^2$ from the quadratic cost. Because persistence falls with order, the removed terms are the most expensive per unit of variation. The truncated rule is a score rather than a set of stocks.

Smooth scores replace thresholds altogether. Cutting a persistent signal at a threshold discards the information that the rank of a stock has moved only slightly, so every small move across the threshold becomes a full trade, whereas a smooth score trades a little whenever the rank moves. When costs are quadratic in the size of a trade, \cref{cor:profile}(ii) makes the expected cost of a score of given variance proportional to one minus its persistence, so the more persistent smooth score is cheaper. When costs are proportional, many small trades can add up, and the comparison is empirical (\cref{sec:resolution}). Two classical results for bivariate normal pairs, Sheppard's formula and the formula for their Spearman correlation \citep{Sheppard1899,Kruskal1958}, measure the loss of persistence.

\begin{proposition}[Thresholds lower persistence]\label{prop:discrete}
Let the latent score of signal $j$ be standard normal, with $(Z_{j,t},Z_{j,t+1})$ bivariate normal with correlation $r_j\in[0,1]$, independent across signals. Let the rank be $u_{j,t}=\Phi(Z_{j,t})$ and the median literal $x_{j,t}=\1\{Z_{j,t}>0\}$. Then
\begin{enumerate}[label=(\roman*),nosep]
\item $\mathrm{corr}(x_{j,t},x_{j,t+1})=\tfrac2\pi\arcsin r_j$;
\item $\mathrm{corr}(u_{j,t},u_{j,t+1})=\tfrac6\pi\arcsin(r_j/2)\ge\tfrac2\pi\arcsin r_j$;
\item the median character on a set $S$ of signals has persistence $\prod_{j\in S}\tfrac2\pi\arcsin r_j$, and the product of centered ranks $\prod_{j\in S}(u_{j,t}-\tfrac12)$ has persistence $\prod_{j\in S}\tfrac6\pi\arcsin(r_j/2)$.
\end{enumerate}
\end{proposition}

\begin{proof}
(i) For a standard bivariate normal pair with correlation $r$, $\Pr[Z_t>0,Z_{t+1}>0]=\tfrac14+\arcsin(r)/(2\pi)$ \citep{Sheppard1899}. The literal is Bernoulli$(\tfrac12)$, so its correlation is $(\Pr[1,1]-\tfrac14)/\tfrac14=\tfrac2\pi\arcsin r$.
(ii) The correlation of $\Phi(Z_t)$ and $\Phi(Z_{t+1})$ is the Spearman correlation of a bivariate normal pair, $\tfrac6\pi\arcsin(r/2)$ \citep{Kruskal1958}. Let $g(r)=3\arcsin(r/2)-\arcsin r$. Then $g(0)=g(1)=0$ and $g''(r)=3r/\bigl(8(1-r^2/4)^{3/2}\bigr)-r/(1-r^2)^{3/2}<0$ on $(0,1)$, so $g\ge0$ on $[0,1]$.
(iii) For independent signals, the expectation of a product of centered functions factors across signals, so the persistence of a product is the product of the persistences.
\end{proof}

With $r=0.95$, a median literal has persistence 0.80 and the rank has 0.95; for a second-order exposure the numbers are 0.64 and 0.89. The loss compounds with order, so under quadratic costs smooth implementations matter most for interactions. \Cref{sec:resolution} tests these predictions.

\section{Data}\label{sec:data}

\subsection{Sample, signals, and timing}

We use CRSP monthly and daily stock files and Compustat annual fundamentals, linked through the CRSP--Compustat Merged database with link types LU and LC and primary links P and C. The universe at the end of month $t$ contains common shares (CRSP share codes 10 and 11) listed on the NYSE, AMEX, or NASDAQ. We keep one security per company, the one with the largest market capitalization, and drop financial firms (SIC codes 6000--6999) and stocks below the 20th percentile of NYSE market capitalization. A stock enters the universe only if all twelve signals are available at $t$. The sample has 623 monthly returns, from February 1973 to December 2024, with 1,257 stocks per month on average. Requiring complete data tilts the universe toward seasoned firms, because beta needs 24 months of returns and long-term reversal needs 36.

\begin{figure}[t]
\centering
\resizebox{\textwidth}{!}{\begin{tikzpicture}[
    font=\small,
    box/.style={draw, rounded corners=2pt, fill=black!4, align=center, inner sep=3pt, font=\scriptsize},
    note/.style={font=\scriptsize, text=black!65, align=center},
    tick/.style={thick},
    yr/.style={font=\scriptsize, text=black!70}
  ]
  \node[anchor=west, font=\small\bfseries] at (-0.2, 2.5) {(a) Monthly timing};
  \draw[thick, ->] (0, 0) -- (14.6, 0);
  \foreach \x in {0.8, 4.0, 8.2, 12.6} \draw[tick] (\x, 0.12) -- (\x, -0.12);
  \node[box, anchor=south] at (0.8, 0.35) {fiscal\\ year end};
  \node[box, anchor=south] at (4.0, 0.35) {accounting data\\ usable (+6 months;\\ dropped after\\ 548 days)};
  \node[box, anchor=south] at (8.2, 0.35) {end of month $t$: ranks,\\ universe filters, literals,\\ rule, base weights $\omega_{i,t}$;\\ trade to target weights};
  \node[box, anchor=south] at (12.6, 0.35) {end of month $t{+}1$:\\ new ranks, rebalance};
  \draw[decorate, decoration={brace, amplitude=5pt, mirror}] (8.3, -0.35) -- (12.5, -0.35)
     node[midway, below=6pt, note] {holding period: delisting-adjusted\\ excess return $r_{i,t+1}$};
  \node[note, anchor=north] at (2.4, -0.35) {Compustat annual};
  \node[note, anchor=north] at (6.1, -0.35) {CRSP prices, returns through\\ month $t$, daily volatility in $t$,\\ 60-month beta};
  \def\yx#1{(#1-1973)*0.27}
  \node[anchor=west, font=\small\bfseries] at (-0.2, -2.75) {(b) Estimation and evaluation windows};
  \begin{scope}[yshift=-4.45cm]
    \draw[thick, ->] (0, 0) -- (14.6, 0);
    \foreach \y in {1973, 1985, 1995, 2005, 2015, 2024} {
      \draw[tick] ({\yx{\y}}, 0.1) -- ({\yx{\y}}, -0.1);
      \node[yr, below] at ({\yx{\y}}, -0.1) {\y};
    }
    \fill[black!18] ({\yx{1973}}, 0.35) rectangle ({\yx{2005}}, 0.65);
    \fill[black!45] ({\yx{2005}}, 0.35) rectangle ({\yx{2024}+0.27}, 0.65);
    \node[note, anchor=south] at ({\yx{1989}}, 0.7) {training window, 1973--2004};
    \node[note, anchor=south] at ({\yx{2014.6}}, 0.7) {evaluation window, 2005--2024};
  \end{scope}
\end{tikzpicture}}
\caption{Timing and estimation windows. Panel (a) shows when information becomes available and when positions are formed and held. Panel (b) shows the split used for the out-of-sample tangency portfolios of \cref{sec:info} and for the rule search of Online Appendix~\ref{oa:search}: weights and selected rules are estimated on 1973--2004 and evaluated on 2005--2024, and inside the training window the ridge penalty is chosen by five-fold cross-validation with contiguous folds.}
\label{fig:timeline}
\end{figure}
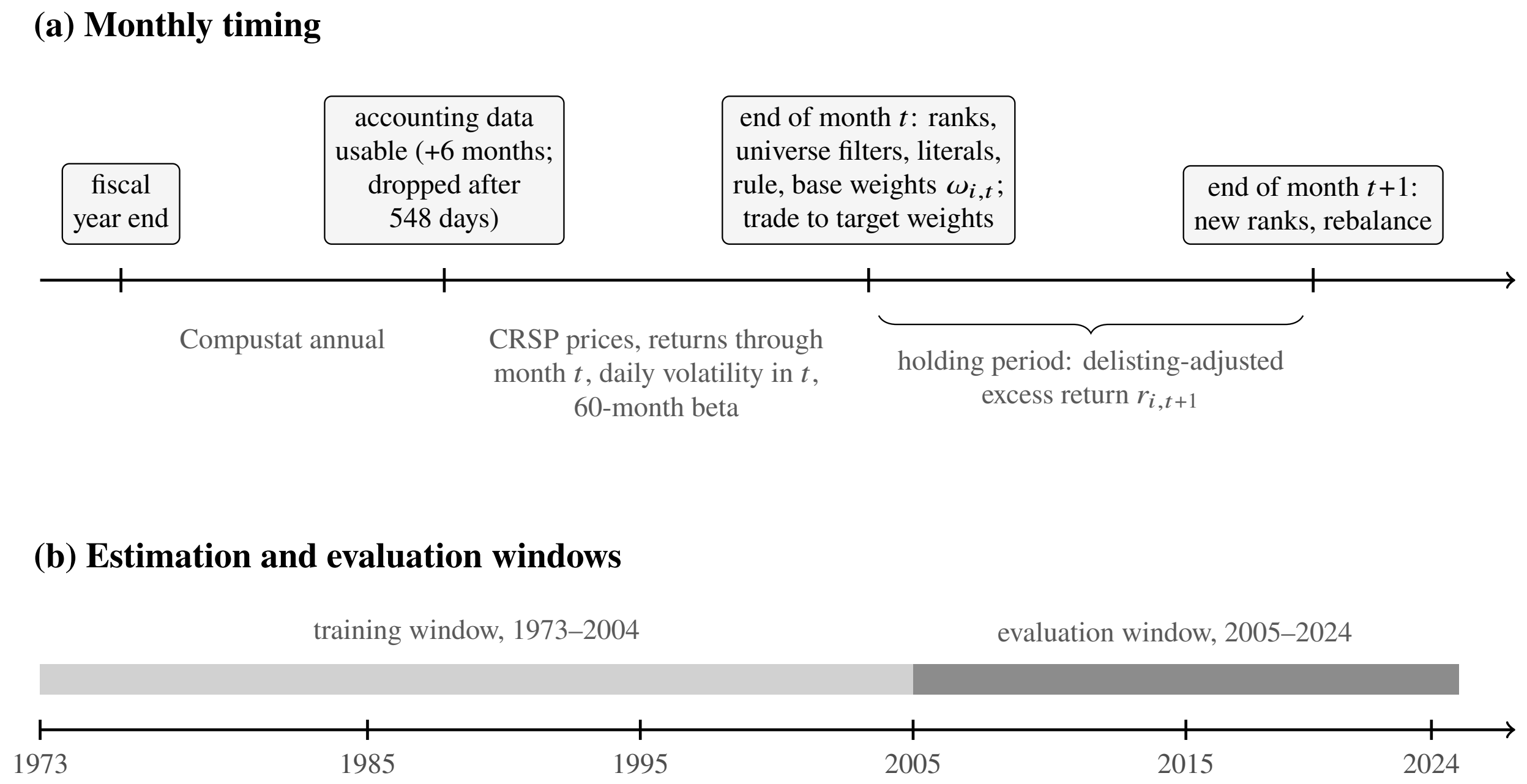

\Cref{fig:timeline}(a) shows when each piece of information becomes available. Accounting variables are used from six months after the fiscal year end and are not carried forward for more than 548 days. Book equity is stockholders' equity, or common equity plus preferred stock, plus deferred taxes and investment tax credit, minus preferred stock measured by redemption, liquidating, or par value; nonpositive book equity is treated as missing. Market variables are measured at the end of month $t$: momentum is the return from month $t-12$ to $t-1$, short-term reversal the return in month $t$, long-term reversal the return from month $t-60$ to $t-12$, volatility the standard deviation of daily returns in month $t$, and beta is estimated from up to 60 monthly returns on the CRSP value-weighted index. The twelve signals, listed in \cref{tab:signals}, are market equity \citep{Banz1981}, book-to-market \citep{RRL1985,FF1992}, momentum \citep{JegadeeshTitman1993}, short-term reversal \citep{Jegadeesh1990}, gross profitability \citep{NovyMarx2013}, return on equity \citep{HaugenBaker1996,HXZ2015}, asset growth \citep{CGS2008}, net share issuance \citep{PontiffWoodgate2008}, volatility \citep{AHXZ2006}, beta \citep{BJS1972,FrazziniPedersen2014}, sales growth \citep{LSV1994}, and long-term reversal \citep{DeBondtThaler1985}. Each signal is ranked within the universe at $t$, with average ranks for ties. Literals, rules, and base weights use only information available at $t$, and positions are held over month $t+1$.

\begin{table}[t]\centering
\caption{Signals, persistence, and first-order portfolios}\label{tab:signals}
\begin{threeparttable}\small\setlength{\tabcolsep}{4pt}
\begin{tabular}{llcccccc}\toprule
 & & & \multicolumn{3}{c}{Persistence} & & \\ \cmidrule(lr){4-6}
 & Definition & Favorable side & Median $\rho_j$ & Three thr. & Rank & Premium (\%/mo) & $t$ \\ \midrule
SIZE & Market equity & small & 0.95 & 0.97 & 0.99 & 0.09 & 1.98 \\
BM & Book-to-market & high & 0.92 & 0.95 & 0.98 & 0.15 & 1.49 \\
MOM & Return, months $t-12$ to $t-1$ & high & 0.73 & 0.84 & 0.90 & 0.21 & 1.77 \\
STR & Return, month $t$ & low & $-$0.02 & $-$0.03 & $-$0.03 & 0.28 & 3.03 \\
GP & Gross profits / assets & high & 0.98 & 0.99 & 0.99 & 0.13 & 1.62 \\
ROE & Net income / book equity & high & 0.96 & 0.96 & 0.97 & 0.03 & 0.53 \\
AG & Asset growth & low & 0.93 & 0.94 & 0.94 & 0.15 & 2.11 \\
ISS & Net share issuance, 12 months & low & 0.90 & 0.92 & 0.94 & 0.13 & 1.61 \\
VOL & Daily return volatility, month $t$ & low & 0.48 & 0.59 & 0.63 & 0.01 & 0.08 \\
BETA & Market beta, 60 months & low & 0.94 & 0.97 & 0.99 & 0.03 & 0.20 \\
SG & Sales growth & low & 0.93 & 0.94 & 0.94 & 0.09 & 1.18 \\
LTR & Return, months $t-60$ to $t-12$ & low & 0.86 & 0.92 & 0.97 & 0.16 & 2.03 \\
\bottomrule\end{tabular}
\begin{tablenotes}\footnotesize
\item Each signal is oriented so that the literal equals one on the side documented to earn higher returns. Persistence is the month-to-month correlation of a stock's score: the sign of the median literal (median $\rho_j$, equal to $1-2\Pr[x_{j,t}\neq x_{j,t+1}]$), the centered mean rank of its bucket at the 20th, 50th, and 80th percentiles (three thr.), and the centered rank $u_{j,i,t}-\tfrac12$ (rank). Premium is the mean of $2\Pi^{\{j\}}_{t+1}=4\beta_t(1-\beta_t)(R^H_{t+1}-R^L_{t+1})$ with capped value weights, where $R^H$ and $R^L$ are the returns of the favorable and unfavorable halves and $\beta_t$ is the base weight of the favorable half; it equals the half-minus-half spread when $\beta_t=1/2$. Sample: 1973--2024, 623 months, 1,257 stocks per month on average.
\end{tablenotes}\end{threeparttable}\end{table}

\subsection{Returns, breakpoints, and weights}

Returns are CRSP monthly returns in excess of the one-month Treasury bill rate and include the CRSP delisting return when the delisting month also has a monthly return. The universe at $t$ does not condition on a return being available at $t+1$. For 0.4\% of stock-months the panel has no return for month $t+1$, and we assign these observations a zero excess return. Nearly all of them are delistings: 89\% are mergers, whose CRSP delisting returns average 2.0\%, and 1\% are performance-related delistings, which tend to have large negative returns \citep{Shumway1997,ShumwayWarther1999}. Replacing the zeros by CRSP delisting returns where available leaves the main results essentially unchanged (\cref{tab:missing}). Breakpoints are percentiles of the universe at $t$: median literals split each signal at the universe median, three-threshold literals at the 20th, 50th, and 80th percentiles, and a robustness version uses the 70th percentile. Base weights are market capitalizations capped at the 80th NYSE percentile, as in \citet{JKP2023}; the cap limits the influence of the largest firms. We call these capped value weights (capped VW in the tables); equal weights (EW) serve as a robustness check.

\subsection{Trading costs}

We proxy one-way trading costs by quoted half-spreads. For each stock and day we compute $\tfrac12(\text{ask}-\text{bid})/\text{midpoint}$ from CRSP closing quotes, drop half-spreads above 25\%, and take the monthly median. CRSP reports closing quotes for all exchanges from the early 1990s, so cost-adjusted results start in 1995. The median half-spread in the universe falls from 76 basis points in 1995 to 2.4 basis points after 2020. Quotes are available for 99.4\% of stock-months after 1995, and the remaining stock-months receive the cross-sectional median of the month. Costs are charged on the change in target weight at each month end, including entries into and exits from the universe, and price drift between rebalancing dates is ignored. When portfolios are combined, trades are netted at the stock level before costs are charged \citep{DMNU2020}. Quoted spreads overstate the cost of small trades in liquid stocks, understate the cost of large trades, and omit short-sale borrowing costs, which we do not model; Online Appendix~\ref{oa:costs} discusses other cost measures.

\subsection{Estimation windows}

Tests that do not estimate portfolio weights, such as the turnover predictions, the cost comparisons, and GRS tests of fixed portfolios, use the full sample, or 1995--2024 when they involve costs; GRS tests of test assets are also reported for samples that start in other years. The out-of-sample tangency portfolios are estimated on 1973--2004 and evaluated on 2005--2024 (\cref{fig:timeline}(b)).

\section{Testing the framework's predictions}\label{sec:costs}

The framework makes three predictions that the data can test: turnover follows the interaction profile and the persistence of the literals, smooth scores are more persistent than thresholds and increasingly so for interactions, and trading costs rise with order.

\subsection{Persistence and turnover}\label{sec:turnover-evidence}

\begin{figure}[t]
\centering
\includegraphics[width=\textwidth]{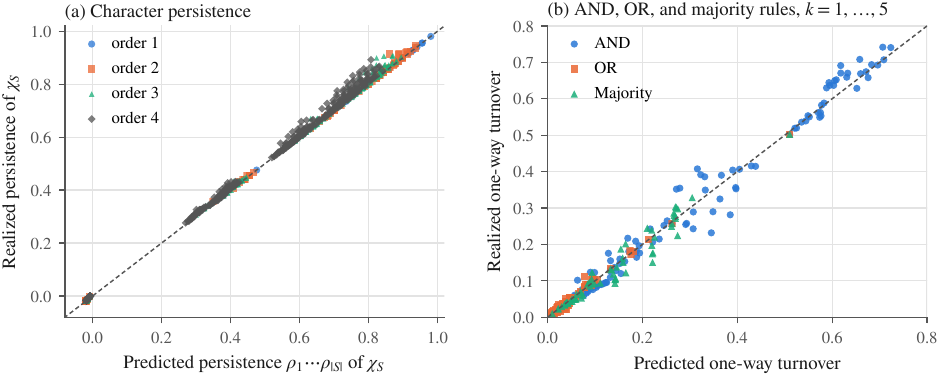}
\caption{Persistence and turnover. Panel (a) plots the realized month-to-month persistence of each median character of order one to four against the product of the persistences of its literals. Panel (b) plots the realized one-way turnover of equal-weighted legs defined by AND, OR, and majority rules on one to five literals against the exact value under \cref{ass:markov}. The dashed line in each panel is the 45-degree line.}
\label{fig:turnover}
\end{figure}

\Cref{prop:noise} predicts that the persistence of a character is the product of the persistences of its literals, and \cref{cor:profile,prop:gates} turn these persistences into the turnover of a rule; both predictions hold closely (\cref{fig:turnover} and \cref{tab:turnover}). Literal persistence ranges from $-0.02$ for short-term reversal, which is redrawn almost every month, to 0.98 for gross profitability (\cref{tab:signals}). Within each order from one to four, the realized persistence of the characters and the product of the persistences of their literals have a correlation of at least 0.999 and a mean absolute difference of at most 0.01. For rules, the predictions fit the realized turnover of 286 equal-weighted AND, OR, and majority legs (262 distinct, because the three rules coincide on a single literal) with a slope of 1.02 and a correlation of 0.99. An intersection of five literals sells almost half of its holdings every month, a single literal sells a tenth, and a union of five literals sells 1.5\%; majority rules fall in between.

Independence across signals is the weakest part of \cref{ass:markov}, and it fails most clearly for accounting signals, whose literals tend to change in the same month because a new annual report updates them together. The flip correlation of two signals is the correlation between the indicators that their literals change from one month to the next. It is 0.38 for asset growth and sales growth, and 0.25 for return on equity and asset growth. Literals that flip together leave their product unchanged, so these interactions are more persistent than predicted. For second-order characters whose signals have a flip correlation of 0.15 or more, realized persistence exceeds the prediction by 0.035 on average, compared with 0.001 when the flip correlation is below 0.05. Legs whose literals include such a pair trade 7\% less than predicted (\cref{tab:independence}). Correlation in levels matters much less: value and return on equity have a level correlation of $-0.37$, yet the persistence of their interaction is within 0.006 of the prediction.

\subsection{Thresholds and smooth scores}\label{sec:resolution}

\Cref{prop:discrete} predicts that the same exposure is more persistent when it is built from a smooth score than from thresholds, and that the advantage compounds in interactions. To test this, we implement each signal at three resolutions: the sign of its median literal, the centered mean rank of its bucket at the 20th, 50th, and 80th percentiles, and its centered rank. An exposure of order $d$ is the product of $d$ such scores, so the three versions target the same exposure with increasing resolution (\cref{sec:walsh}).

\begin{table}[t]\centering
\caption{Persistence and turnover of the same exposures at three resolutions}\label{tab:resolution}
\begin{threeparttable}\footnotesize\setlength{\tabcolsep}{4pt}
\begin{tabular}{cccccccccc}\toprule
\multicolumn{10}{l}{\emph{Panel A. Persistence}}\\
 & & \multicolumn{3}{c}{Realized} & \multicolumn{3}{c}{Product of persistences} & & \\ \cmidrule(lr){3-5}\cmidrule(lr){6-8}
Order & $N$ & Median & Three thr. & Rank & Median & Three thr. & Rank & Correlation & Ordered \\ \midrule
1 & 12 & 0.796 & 0.829 & 0.853 & -- & -- & -- & -- & 11 \\
2 & 66 & 0.625 & 0.685 & 0.725 & 0.626 & 0.680 & 0.720 & 0.999 & 57 \\
3 & 220 & 0.493 & 0.575 & 0.624 & 0.486 & 0.551 & 0.601 & 0.998 & 195 \\
\bottomrule\end{tabular}

\medskip
\begin{tabular}{ccccccccc}\toprule
\multicolumn{9}{l}{\emph{Panel B. Turnover under proportional costs}}\\
 & & \multicolumn{4}{c}{Per unit of risk} & \multicolumn{3}{c}{Per unit of gross exposure} \\ \cmidrule(lr){3-6}\cmidrule(lr){7-9}
Order & $N$ & Median & Three thr. & Rank & Rank lower & Median & Three thr. & Rank \\ \midrule
1 & 12 & 0.205 & 0.232 & 0.241 & 0 & 0.205 & 0.258 & 0.279 \\
2 & 66 & 0.398 & 0.395 & 0.373 & 61 & 0.403 & 0.499 & 0.494 \\
3 & 220 & 0.524 & 0.463 & 0.421 & 220 & 0.525 & 0.654 & 0.661 \\
\bottomrule\end{tabular}
\begin{tablenotes}\footnotesize
\item Each signal enters as the sign of its median literal (median), the centered mean rank of its bucket at the 20th, 50th, and 80th percentiles (three thr.), or its centered rank (rank); an exposure of order $d$ is the product of $d$ such scores, one for each of $d$ signals, so the three versions target the same exposure with increasing resolution. Realized persistence is the correlation of a stock's score in months $t$ and $t+1$, pooled over 1973--2024 and averaged over the $N$ sets of signals; for median literals it differs slightly from the mean product $\E[\chi_S(\mathbf x_t)\chi_S(\mathbf x_{t+1})]$ of Tables~\ref{tab:turnover} and~\ref{tab:independence} because the products do not have exactly zero mean. Product of persistences is the average prediction of \cref{prop:noise,prop:discrete}, the product of the first-order persistences of the signals involved. Correlation is the smallest, across the three resolutions, of the correlation between realized and predicted persistence across sets of signals. Ordered is the number of sets whose persistence rises from median to three thresholds to rank. Panel B reports the monthly sum of absolute weight changes of continuing stocks in the equal-weighted portfolio with weights proportional to the demeaned score, averaged over months and sets of signals. Per unit of risk divides it by $\sqrt{n\sum_ia_i^2}$, which compares scores with the same cross-sectional variance; per unit of gross exposure divides it by $\sum_i|a_i|$. Rank lower is the number of sets for which the rank version turns over less than the median version per unit of risk.
\end{tablenotes}\end{threeparttable}\end{table}

\Cref{tab:resolution} confirms both predictions. At first order, mean persistence rises from 0.796 for median literals to 0.829 for three thresholds and 0.853 for ranks, and the ordering holds for 11 of the 12 signals; the exception is short-term reversal, which is redrawn almost every month at any resolution (\cref{tab:signals}). At second order, mean persistence is 0.63, 0.69, and 0.73, and the ordering holds for 57 of the 66 pairs; at third order it is 0.49, 0.58, and 0.62, and it holds for 195 of the 220 triples. The gap between ranks and median literals widens from 0.06 at first order to 0.10 and 0.13 at second and third order, the compounding that \cref{prop:discrete} implies. The product rule also holds at every resolution: averaged over pairs, the products of first-order persistences are 0.63, 0.68, and 0.72, and across pairs and triples the correlation between realized and predicted persistence is at least 0.998. The cost implications depend on the scale at which implementations are compared and on the cost function. We compare them at equal risk, with scores scaled to the same cross-sectional variance, which is the scale of the exposures in the decomposition. Under quadratic costs, the expected cost is then proportional to one minus persistence (\cref{cor:profile}(ii)), so rank-product interactions are about a quarter cheaper to trade than median interactions of the same order. Under proportional costs, a smooth score trades a little whenever ranks move, and at first order this outweighs its higher persistence: monthly turnover per unit of risk is 0.241 with ranks and 0.205 with median literals (Panel B of \cref{tab:resolution}). For interactions the compounding loss of persistence dominates. Turnover per unit of risk is 0.373 with rank products and 0.398 with median literals at second order, lower for 61 of the 66 pairs, and 0.421 against 0.524 at third order, lower for all 220 triples. Per unit of gross exposure, which matters under leverage limits, smooth scores turn over more at every order.

\subsection{Trading costs by order}\label{sec:cost-evidence}

Across the 298 median portfolios of orders one to three, turnover has a correlation of 0.999 with $1-\rho_S$, and because costs are charged on trades, they rise with order. \Cref{tab:costs} charges quoted half-spreads to every portfolio over 1995--2024, and \cref{fig:costs} plots the averages. Per unit of gross exposure, the mean absolute premium falls from 5.4 basis points per month at first order to 3.5 and 2.6 at second and third order, while the mean cost rises from 4.8 to 7.5 and 9.5. As a result, 58\% of first-order portfolios have absolute premia above their quoted-spread costs, against 24\% at second order and 6\% at third order.

The pattern also appears with equal weights, under which every median literal has coverage one half and each second-order portfolio is a pure interaction contrast: the shares are 75\%, 17\%, and 2\% (\cref{tab:costsens}).

\begin{table}[t]\centering
\caption{Premia and trading costs by order}\label{tab:costs}
\begin{threeparttable}\small
\begin{tabular}{lcccccc}\toprule
\multicolumn{7}{l}{\emph{Panel A. Individual portfolios by order, 1995--2024}}\\
Order & $N$ & $1-\rho_S$ & Turnover & $|$Premium$|$ (bp) & Cost (bp) & $|$Premium$|>$ cost (\%) \\ \midrule
1 & 12 & 0.204 & 0.264 & 5.42 & 4.82 & 58.3 \\
2 & 66 & 0.371 & 0.432 & 3.47 & 7.51 & 24.2 \\
3 & 220 & 0.507 & 0.551 & 2.56 & 9.47 & 5.9 \\
\midrule
\multicolumn{7}{l}{\emph{Panel B. Tangency portfolios up to a given order, 2005--2024}}\\
Order & Turnover & Gross (bp) & Cost (bp) & SR gross & SR net & \\ \midrule
$\le 1$ & 0.76 & 4.52 & 1.83 & 0.18 & 0.11 & \\
$\le 2$ & 0.96 & 7.75 & 2.71 & 0.48 & 0.32 & \\
$\le 3$ & 0.96 & 8.80 & 2.28 & 0.61 & 0.45 & \\
\bottomrule\end{tabular}
\begin{tablenotes}\footnotesize
\item Premium and cost are monthly, per unit of gross exposure, with quoted half-spreads applied to all trades. Turnover is two-sided per unit of gross exposure. In Panel B the weights are ridge estimates from 1973--2004 and are held fixed out of sample. Sharpe ratios are annualized. Across the 298 portfolios of Panel A, turnover has a correlation of 0.999 with $1-\rho_S$.
\end{tablenotes}\end{threeparttable}\end{table}

\section{Interaction exposures and expected returns}\label{sec:economics}

By \cref{cor:span}, the character portfolios of order at most $k$ span every rule on $k$ signals at the chosen thresholds, so tests on them cover all such rules at once. This section uses them to ask whether interaction exposures carry information about expected returns beyond first-order exposures and whether they separate factor models better than conventional sorts.

\subsection{Information in interaction portfolios}\label{sec:info}

\Cref{ex:ff} gives a direct illustration in a standard factor construction, and \cref{tab:ff2222} applies it to our universe. The construction differs from the $2\times2\times2\times2$ factors of \citet{FF2015} in using gross profitability, universe medians, and capped value weights, so the four first-order coordinates are analogues of those factors. The benchmarks in our tests, the five-factor model (FF5) and FF6, use the standard $2\times3$ factors from Kenneth French's data library. Three of the eleven interaction coordinates have $|t|>2$, and all three remain significant against FF5. The largest is the third-order size--value--investment coordinate, $-0.12\%$ per month with a $t$-statistic of $-4.1$ and an FF5 alpha $t$-statistic of $-5.5$. The interaction between value and investment thus differs between small and large stocks.

\begin{table}[t]\centering
\caption{First-order coordinates and interactions of the $2\times2\times2\times2$ sort}\label{tab:ff2222}
\begin{threeparttable}\small
\begin{tabular}{lccccc}\toprule
Coordinate & Order & Mean (\%/mo) & $t$ & FF5 $\alpha$ (\%/mo) & $t(\alpha)$ \\ \midrule
Prof.$^{\ast}$ & 1 & 0.23 & 3.40 & 0.17 & 2.55 \\
Inv. & 1 & 0.16 & 2.94 & 0.06 & 1.59 \\
Size & 1 & 0.15 & 1.76 & 0.09 & 2.26 \\
Value & 1 & 0.22 & 2.18 & $-$0.02 & $-$0.29 \\
\addlinespace
Prof.$^{\ast}$$\times$Inv. & 2 & $-$0.09 & $-$2.84 & $-$0.09 & $-$2.87 \\
Size$\times$Prof.$^{\ast}$ & 2 & 0.04 & 1.39 & 0.03 & 0.86 \\
Size$\times$Inv. & 2 & 0.03 & 1.05 & 0.06 & 1.98 \\
Size$\times$Value & 2 & 0.07 & 1.91 & 0.04 & 1.11 \\
Value$\times$Prof.$^{\ast}$ & 2 & $-$0.01 & $-$0.20 & $-$0.04 & $-$1.09 \\
Value$\times$Inv. & 2 & $-$0.09 & $-$2.61 & $-$0.09 & $-$2.58 \\
\addlinespace
Size$\times$Prof.$^{\ast}$$\times$Inv. & 3 & $-$0.05 & $-$1.79 & $-$0.05 & $-$1.52 \\
Size$\times$Value$\times$Prof.$^{\ast}$ & 3 & $-$0.02 & $-$0.76 & 0.00 & $-$0.07 \\
Size$\times$Value$\times$Inv. & 3 & $-$0.12 & $-$4.06 & $-$0.16 & $-$5.46 \\
Value$\times$Prof.$^{\ast}$$\times$Inv. & 3 & 0.02 & 0.57 & 0.03 & 0.89 \\
\addlinespace
Size$\times$Value$\times$Prof.$^{\ast}$$\times$Inv. & 4 & 0.01 & 0.17 & $-$0.01 & $-$0.39 \\
\bottomrule\end{tabular}
\begin{tablenotes}\footnotesize
\item The sixteen cells are intersections of median literals on size, book-to-market, gross profitability ($^{\ast}$used in place of operating profitability), and asset growth, with capped value weights inside each cell. Each row reports $2\hat g(A)$, twice the Walsh coefficient of the cell-return function on the set $A$. The four first-order coefficients are analogues of SMB, HML, RMW, and CMA; they differ from the published factors in universe, breakpoints, weights, and the profitability measure. The eleven higher-order coefficients are interaction spreads that the four factors leave out. FF5 $\alpha$ is the intercept from a regression on the five Fama--French factors. Sample: 1973--2024.
\end{tablenotes}\end{threeparttable}\end{table}

Interactions also carry information when all twelve signals are used. Over the full sample, block GRS tests reject zero intercepts of the second-order portfolios on all first-order ones ($p<0.001$) and of the third-order portfolios on all portfolios of order at most two ($p=0.006$; \cref{tab:blocks}). Out of sample, with weights estimated on 1973--2004, the tangency portfolio of the twelve first-order portfolios earns a Sharpe ratio of 0.18 over 2005--2024. Adding the 66 second-order portfolios raises it to 0.48, and the alpha of the second tangency portfolio on the first has a Newey--West $t$-statistic of 2.5 (six lags; \citealp{NeweyWest1987}). With literals at the 70th percentile, the two Sharpe ratios are 0.13 and 0.64 and the $t$-statistic is 3.3. Third-order portfolios add less: they raise the Sharpe ratio to 0.61 with median literals ($t=1.8$) and lower it to 0.41 with 70th-percentile literals. The gain from second-order portfolios comes from 2005--2019, when the Sharpe ratio of the first-order tangency portfolio is 0.04 and that of order at most two 0.48; over 2020--2024 both earn about 0.5 (\cref{fig:oosdegree}). After quoted-spread costs, the out-of-sample Sharpe ratios are 0.11 for first-order portfolios, 0.32 with second-order portfolios added, and 0.45 with third-order portfolios added (Panel B of \cref{tab:costs}). These portfolios cost less per unit of exposure than the individual portfolios of \cref{sec:cost-evidence}, mainly because spreads were narrower after 2005.

\subsection{Interaction portfolios as test assets}\label{sec:miss}

By \cref{cor:span}, the median portfolios of order at most two span the managed portfolio of every rule built on the median literals of two signals. Conventional sorts use other breakpoints and coverage-normalized weights, so they lie outside this span. \Cref{tab:testassets} compares thirteen sets of test assets built from the same signals, universe, and weights. Five are conventional: 120 deciles, 275 size-by-signal $5\times5$ portfolios, 594 portfolios from $3\times3$ sorts on every pair of signals, 12 decile spreads, and 132 corner spreads from the pairwise $3\times3$ sorts. The other eight are median and three-threshold portfolios and their smooth counterparts, the linear rank and rank-product portfolios.

\begin{table}[t]\centering
\caption{Test assets built from the same twelve signals}\label{tab:testassets}
\begin{threeparttable}\footnotesize\setlength{\tabcolsep}{4pt}
\begin{tabular}{lccccccc}\toprule
 & & \multicolumn{2}{c}{FF6, 1973--2024} & \multicolumn{2}{c}{FF6, 1995--2024} & \multicolumn{2}{c}{$q^5$, 1995--2019} \\ \cmidrule(lr){3-4}\cmidrule(lr){5-6}\cmidrule(lr){7-8}
Test assets & $N$ & $|t|>2$ (\%) & GRS $p$ & $|t|>2$ (\%) & GRS $p$ & $|t|>2$ (\%) & GRS $p$ \\ \midrule
Deciles & 120 & 13.3 & $<$0.001 & 5.0 & 0.260 & 2.5 & 0.431 \\
Size$\times$signal $5\times5$ & 275 & 16.6 & 0.001 & 2.5 & 0.895 & 4.4 & 0.673 \\
Pairwise $3\times3$ & 594 & 16.8 & 0.034 & 5.2 & -- & 3.7 & -- \\
Decile spreads & 12 & 25.0 & 0.001 & 16.7 & 0.258 & 0.0 & 0.470 \\
$3\times3$ corner spreads & 132 & 26.5 & $<$0.001 & 7.6 & 0.241 & 1.5 & 0.138 \\
\addlinespace
Linear rank portfolios & 12 & 25.0 & $<$0.001 & 8.3 & 0.451 & 0.0 & 0.592 \\
\addlinespace
Median portfolios, order 1 & 12 & 25.0 & $<$0.001 & 8.3 & 0.362 & 0.0 & 0.518 \\
Median portfolios, order 2 & 66 & 28.8 & $<$0.001 & 12.1 & 0.086 & 7.6 & 0.144 \\
Median portfolios, order $\le2$ & 78 & 28.2 & $<$0.001 & 11.5 & 0.007 & 6.4 & 0.037 \\
Median portfolios, order 3 & 220 & 17.7 & $<$0.001 & 9.5 & 0.003 & 5.9 & 0.090 \\
Three-threshold portfolios, order 1 & 36 & 19.4 & $<$0.001 & 11.1 & 0.372 & 0.0 & 0.761 \\
Three-threshold portfolios, order 2 & 594 & 17.3 & 0.038 & 8.4 & -- & 6.4 & -- \\
Rank-product portfolios & 66 & 24.2 & $<$0.001 & 13.6 & 0.005 & 4.5 & 0.014 \\
\bottomrule\end{tabular}
\begin{tablenotes}\footnotesize
\item All test assets are built from the same twelve signals, universe, and capped value weights. Sorts are conventional portfolios; median and three-threshold portfolios are the zero-cost character portfolios $\Pi^S$ of Section~\ref{sec:walsh}, and rank portfolios use linear ranks and rank products. GRS $p$-values are omitted when $N$ is too large relative to $T$. The 1995--2024 window starts when quoted spreads become available for all exchanges (\cref{sec:data}); $q^5$ factors are available to us through 2019.
\end{tablenotes}\end{threeparttable}\end{table}

Over the full sample every set rejects FF6 at the 5\% level, so the test does not separate them. In more recent samples it does (\cref{tab:windows}): in samples that start in 1995, 2000, or 2005, no conventional set rejects FF6, whereas the median portfolios of order at most two do ($p$ between 0.004 and 0.010). Over 1995--2024 (through 2019 for $q^5$), the GRS $p$-values of the conventional sets range from 0.24 to 0.90 for FF6 and from 0.14 to 0.67 for $q^5$, and 5.2\% of the 594 pairwise portfolios have $|t|>2$ against FF6, close to the share expected under the null. Over the same period, FF6 is rejected by the median portfolios of order at most two and of order three ($p=0.007$ and $0.003$) and by the rank-product portfolios ($p=0.005$), and $q^5$ by the median portfolios of order at most two and the rank-product portfolios ($p=0.037$ and $0.014$). Neither block of median portfolios rejects FF6 on its own: the 12 first-order portfolios give $p=0.36$ and the 66 second-order portfolios $p=0.086$. Corner spreads do not isolate interactions. The high--high minus low--low spread equals the sum of the two first-order contrasts, and the high--low minus low--high spread equals their difference; the interaction contrast $R^{HH}+R^{LL}-R^{HL}-R^{LH}$ is orthogonal to both. Corner spreads therefore contain no interaction contrast, whereas the decomposition supplies it alongside the first-order contrasts, and the two blocks reject together.

\section{Implications for portfolio construction}\label{sec:design}

The framework lets a portfolio manager evaluate a rule before trading it, and the evidence in \cref{sec:costs,sec:economics} shows where the trade-off between interaction exposures and their costs lies. Three steps summarize the implications.

The first step is to measure what a rule holds and what it will cost. The decomposition gives the interaction profile of the rule, and \cref{cor:profile} turns the profile and the persistence of its literals into expected turnover. With literal persistence of 0.9, a second-order exposure turns over about twice as fast as a first-order one, and a third-order exposure nearly three times as fast. Multiplied by a per-trade cost, these turnovers give the cost of each exposure before any trade is placed.

The second step is to weigh premia against costs. Few interaction portfolios cover their own quoted-spread costs over 1995--2024, 24\% at second order and 6\% at third order, but in the out-of-sample tangency portfolios of \cref{sec:info}, evaluated over 2005--2024 when spreads were narrower, the net Sharpe ratio is 0.32 with second-order portfolios and 0.11 without them (\cref{tab:costs}). Third-order portfolios raise the Sharpe ratio further with median literals (from 0.48 to 0.61 gross and from 0.32 to 0.45 net) but lower it with 70th-percentile literals (from 0.64 to 0.41 gross), so stopping at second order, or truncating higher orders (\cref{sec:levers}), is the more robust choice.

The third step is to choose the implementation. At equal risk, smooth scores implement interactions more cheaply than thresholds: about a quarter more cheaply under quadratic costs, as in models of price impact \citep{GarleanuPedersen2013}, and by 6\% at second order and 20\% at third order under proportional costs (\cref{sec:resolution}). For first-order exposures, and under leverage limits, thresholds trade less under proportional costs. Buffer zones are a further lever: they raise the persistence of each literal without adding small trades, and because persistence multiplies, the framework predicts that they help interactions most (\cref{sec:levers}).

\section{Conclusion}\label{sec:conclusion}

Many common portfolio rules are Boolean functions of threshold statements about signal ranks, and the Walsh decomposition describes both the exposures they carry and the turnover they generate. Because the persistence of an interaction is the product of the persistences of its components, rules that load on higher-order interactions trade more, and because thresholds discard information, threshold implementations of an interaction are less persistent than smooth ones and, at equal risk, costlier to trade.

In U.S.\ stocks, these predictions hold: predicted and realized persistence and turnover line up closely, the same interactions are more persistent as rank products than as threshold rules, and trading costs rise with order. The interaction exposures that the decomposition isolates carry information about expected returns beyond first-order exposures, and over 1995--2024, used as test assets, they reveal pricing errors in a standard factor model that conventional sorts miss.

These results suggest treating rule design as part of factor design. Thresholds, buffers, and the order of interactions can be chosen to maximize net rather than gross performance, and the decision rules behind published factors and implemented strategies can be audited to separate intended exposures from those carried by construction; Online Appendices~\ref{oa:nse} and~\ref{oa:verify} give two examples.

\bibliographystyle{chicago}
\bibliography{refs}

\clearpage
\setcounter{page}{1}
\renewcommand{\thepage}{OA-\arabic{page}}
\setcounter{section}{0}
\renewcommand{\thesection}{OA.\arabic{section}}
\setcounter{table}{0}
\renewcommand{\thetable}{OA.\arabic{table}}
\setcounter{figure}{0}
\renewcommand{\thefigure}{OA.\arabic{figure}}
\setcounter{equation}{0}
\renewcommand{\theequation}{OA.\arabic{equation}}
\begin{center}
{\Large Online Appendix}\\[6pt]
{\large The Logic of Factor Portfolios: Boolean Rules, Interactions, and Trading Costs}
\end{center}
\suppressfloats[t]

\section{Proofs and derivations}\label{oa:proofs}

\paragraph{Character portfolios.}
For a first-order portfolio, let $H$ be the stocks with $x_j=1$, $\beta_t$ their base weight, and $R^H_{t+1}$ and $R^L_{t+1}$ the base-weighted average returns of the stocks with $x_j=1$ and $x_j=0$. Since $\chi_{\{j\}}=2x_j-1$ and $\sum_i\omega_{i,t}\tilde r_{i,t+1}=0$,
\[
\Pi^{\{j\}}_{t+1}=2\sum_{i\in H}\omega_{i,t}\tilde r_{i,t+1}=2\beta_t\bigl(R^H_{t+1}-\bar r_{t+1}\bigr)=2\beta_t(1-\beta_t)\bigl(R^H_{t+1}-R^L_{t+1}\bigr),
\]
where the last step uses $\bar r_{t+1}=\beta_tR^H_{t+1}+(1-\beta_t)R^L_{t+1}$. For a second-order portfolio, the literals of signals $j$ and $k$ split the universe into four cells $c\in\{HH,HL,LH,LL\}$, where the first letter refers to signal $j$. With cell weights $\beta^c_t$ and cell returns $R^c_{t+1}$, the same calculation gives $\Pi^{\{j,k\}}_{t+1}=\sum_c\beta^c_t\,\chi_{\{j,k\}}(c)\bigl(R^c_{t+1}-\bar r_{t+1}\bigr)$. When the four cells have equal weight, the signs $\chi_{\{j,k\}}(c)$ sum to zero and the expression in \cref{sec:walsh} follows.

Suppose instead that each of the two literals has coverage one half, as with median literals and equal weights, but the cells may have unequal weight because the signals are correlated. Then $\beta^{HH}_t=\beta^{LL}_t=a$ and $\beta^{HL}_t=\beta^{LH}_t=b$ with $a+b=\tfrac12$, and substituting $\bar r_{t+1}=a(R^{HH}+R^{LL})+b(R^{HL}+R^{LH})$ gives
\[
\Pi^{\{j,k\}}_{t+1}=4ab\bigl[(R^{HH}_{t+1}+R^{LL}_{t+1})-(R^{HL}_{t+1}+R^{LH}_{t+1})\bigr],
\]
a pure interaction contrast. When a literal has coverage $\beta_c\neq\tfrac12$, the portfolio on $T\cup\{c\}$ contains approximately $(2\beta_c-1)\Pi^T$, a scaled copy of the lower-order portfolio, which a regression on lower-order portfolios removes. For third and higher orders, correlation between signals can create such components even when every literal has coverage one half.

\paragraph{Proof of \cref{cor:span}.}
(i) Suppose $f$ depends only on the literals in a set $T$ with $|T|\le k$, and let $S\not\subseteq T$. Pick $j\in S\setminus T$ and let $\mathbf z^{(j)}$ be $\mathbf z$ with coordinate $j$ flipped. Then $f(\mathbf z^{(j)})=f(\mathbf z)$ and $\chi_S(\mathbf z^{(j)})=-\chi_S(\mathbf z)$. Because $\mathbf z\mapsto\mathbf z^{(j)}$ is a bijection of $\{0,1\}^n$,
\[
\sum_{\mathbf z}f(\mathbf z)\chi_S(\mathbf z)=\sum_{\mathbf z}f(\mathbf z^{(j)})\chi_S(\mathbf z^{(j)})=-\sum_{\mathbf z}f(\mathbf z)\chi_S(\mathbf z),
\]
so $\hat f(S)=0$. By \cref{prop:spectral}, $F^f$ is a combination of the $\Pi^S$ with $\emptyset\neq S\subseteq T$, all of order at most $k$. Conversely, each $\chi_S$ with $|S|\le k$ belongs to $\F_k$, and $\Pi^S=F^{\chi_S}$.
(ii) Multiply the regression for $\Pi^S$ by $\hat f(S)$ and sum over $S\neq\emptyset$. By \cref{prop:spectral} the left side is $F^f$, and the intercept of the sum is $\sum_{S\neq\emptyset}\hat f(S)\alpha_S$.
(iii) A portfolio in the span has returns $\mathbf w'\bm\Pi_t$ for some $\mathbf w\neq\mathbf 0$, intercept $\alpha_P=\mathbf w'\bm\alpha$, and residual variance $\sigma^2(\varepsilon_P)=\mathbf w'\Sigma_\varepsilon\mathbf w$. With $\mathbf u=\Sigma_\varepsilon^{1/2}\mathbf w$ and $\mathbf v=\Sigma_\varepsilon^{-1/2}\bm\alpha$, the Cauchy--Schwarz inequality gives $(\mathbf w'\bm\alpha)^2=(\mathbf u'\mathbf v)^2\le(\mathbf u'\mathbf u)(\mathbf v'\mathbf v)=(\mathbf w'\Sigma_\varepsilon\mathbf w)(\bm\alpha'\Sigma_\varepsilon^{-1}\bm\alpha)$, with equality if and only if $\mathbf u\propto\mathbf v$, that is, $\mathbf w\propto\Sigma_\varepsilon^{-1}\bm\alpha$. With $N$ test assets, $N_G$ factors, and factor means $\hat{\bm\mu}_G$ and covariance $\hat\Omega_G$, the GRS statistic $\frac{T-N-N_G}{N}\,\hat{\bm\alpha}'\hat\Sigma_\varepsilon^{-1}\hat{\bm\alpha}/(1+\hat{\bm\mu}_G'\hat\Omega_G^{-1}\hat{\bm\mu}_G)$ is a scaled sample version of the bound. \qed

\paragraph{The $2\times2\times2\times2$ sort.}
In \cref{ex:ff}, $\chi_{\{h\}}(\mathbf z)=\pm1$ according to $z_h$, so $2\hat g(\{h\})=\tfrac18\sum_{\mathbf z:z_h=1}g(\mathbf z)-\tfrac18\sum_{\mathbf z:z_h=0}g(\mathbf z)$. Let $D_r$ be the $h$ spread among the eight cells with $z_{h'}=r$. Because $\chi_{\{h,h'\}}(\mathbf z)$ is $+1$ on the cells where $z_h=z_{h'}$ and $-1$ elsewhere, $\sum_{\mathbf z}g(\mathbf z)\chi_{\{h,h'\}}(\mathbf z)=4D_1-4D_0$, so $2\hat g(\{h,h'\})=\tfrac12(D_1-D_0)$. The coefficients equal the character portfolios when the base weights give every cell the same total weight, for example $\omega'_{i,t}=\omega_{i,t}/\bigl(16\,\beta_t(\mathbf z_{i,t})\bigr)$ with $\beta_t(\mathbf z)$ the base weight of cell $\mathbf z$. Under these weights each cell contributes $g(\mathbf z)/16$ to $\sum_i\omega'_{i,t}r_{i,t+1}$, so $\Pi^A=\tfrac1{16}\sum_{\mathbf z}\chi_A(\mathbf z)\bigl(g(\mathbf z)-\bar r'\bigr)=\hat g(A)$ for $A\neq\emptyset$.

\paragraph{Three-threshold characters.}
With thresholds at the 20th, 50th, and 80th percentiles, the four buckets from bottom to top have probabilities $(0.2,0.3,0.3,0.2)$. Gram--Schmidt in the sequence median, top, bottom gives, listing values on buckets $0$ to $3$,
\[
\psi_{j,1}=(-1,-1,1,1),\qquad \psi_{j,2}=\bigl(0,0,-\tfrac{2}{\sqrt3},\sqrt3\bigr),\qquad \psi_{j,3}=\bigl(-\sqrt3,\tfrac{2}{\sqrt3},0,0\bigr).
\]
Each has mean zero and variance one, and the three are uncorrelated; for example, $\E[\psi_{j,1}\psi_{j,2}]=0.3\cdot(-\tfrac2{\sqrt3})+0.2\cdot\sqrt3=0$. With multi-indices, the proof of \cref{prop:spectral} is unchanged once $\chi_S$ is replaced by $\psi_{\bm a}$ and $\psi_{\bm 0}\equiv1$. For \cref{cor:span}(i), suppose $f$ depends only on the signals in $T$ and $a_j\neq0$ for some $j\notin T$. Under the product measure $\pi=\otimes_l\pi_l$, $f\prod_{l\neq j}\psi_{l,a_l}$ does not depend on $\kappa_j$, so $\hat f(\bm a)=\E_\pi\bigl[f\prod_{l\neq j}\psi_{l,a_l}\bigr]\E_{\pi_j}[\psi_{j,a_j}]=0$, because $\psi_{j,a_j}$ is orthogonal to $\psi_{j,0}\equiv1$. The grouping of terms by the set of signals $\{j:a_j\neq0\}$ does not depend on the sequence of the Gram--Schmidt steps. Finally, if each rank is replaced by the mean rank of its bucket, $\bar u_j(\kappa_j)$, the function $\bar u_j(\kappa_j)-\tfrac12$ has mean zero under $\pi_j$, so $\bigl(\bar u_j(\kappa_j)-\tfrac12\bigr)\bigl(\bar u_k(\kappa_k)-\tfrac12\bigr)$ is a combination of second-order three-threshold characters. Because $|u_j-\bar u_j(\kappa_j)|$ is at most the width $w$ of the widest bucket and centered ranks lie in $[-\tfrac12,\tfrac12]$, this product differs from the rank product by at most $w$.

\paragraph{Proof of \cref{prop:gates}.}
Conditional on $\x_t$, the literals at $t+1$ are independent across $j$, and each moves according to $a_j=\bigl(p_j^2+\rho_jp_j(1-p_j)\bigr)/p_j$ and $b_j=p_j(1-p_j)(1-\rho_j)/(1-p_j)$.
(i) A stock is in the intersection at $t$ only if all its literals equal one, and it stays only if all of them stay at one, which has probability $\prod_{j\in G}a_j$.
(ii) Let $O_t=\mathrm{OR}_G(\x_t)$. Because the union is off exactly when all its literals are zero,
\[
\Pr[O_t=1,O_{t+1}=0]=\Pr[O_{t+1}=0]-\Pr[O_t=0,O_{t+1}=0]=\prod_{j\in G}(1-p_j)-\prod_{j\in G}(1-p_j)(1-b_j).
\]
Dividing by $\Pr[O_t=1]=1-\prod_{j\in G}(1-p_j)$ gives (ii).
(iii) By (i), $\tau(\mathrm{AND}_{G\cup\{l\}})=1-a_l\bigl(1-\tau(\mathrm{AND}_G)\bigr)\ge\tau(\mathrm{AND}_G)$ because $a_l\le1$. For the union, let $q=1-p\in(0,1)$ and $s=1-b\in[0,1]$. With common parameters, (ii) reads $\tau_k=q^k(1-s^k)/(1-q^k)$ for a union of $k$ literals. If $b=0$, every $\tau_k$ is zero. Otherwise $\tau_k>0$. Let $P_k(x)=\sum_{i=0}^{k-1}x^i$, so that $1-s^k=(1-s)P_k(s)$ and $1-q^k=(1-q)P_k(q)$. Then
\[
\frac{\tau_{k+1}}{\tau_k}=q\,\frac{1-s^{k+1}}{1-s^k}\,\frac{1-q^k}{1-q^{k+1}}=\frac{q\,h_k(s)}{h_k(q)},\qquad h_k(x)=\frac{P_{k+1}(x)}{P_k(x)}=1+\frac{x^k}{P_k(x)}.
\]
So $\tau_{k+1}\le\tau_k$ if and only if $q\,h_k(s)\le h_k(q)$. For $x\in(0,1]$, $x^k/P_k(x)=1/\sum_{i=0}^{k-1}x^{i-k}$, and each $x^{i-k}$ with $i<k$ is decreasing in $x$, so $h_k$ is non-decreasing on $[0,1]$ and $h_k(s)\le h_k(1)=(k+1)/k$. Moreover, $q(k+1)/k\le h_k(q)$ is equivalent to $(k+1)\sum_{i=1}^kq^i\le k+k\sum_{i=1}^kq^i$, that is, to $\sum_{i=1}^kq^i\le k$, which holds because $q<1$. Hence $\tau_{k+1}\le\tau_k$. Finally, $\tau_1=q\,b/p=(1-\rho)(1-p)$, the turnover $1-a$ of a single literal. \qed

\paragraph{Threshold rules.}
If $p_j=p$ and $\rho_j=\rho$ for all $j\in G$, the threshold rule $T_c=\1\{\sum_{j\in G}x_j\ge c\}$, $1\le c\le k$, has
\[
\tau(T_c)=\frac{\sum_{m=c}^{k}\binom km p^m(1-p)^{k-m}\Pr[B_m+B'_{k-m}<c]}{\sum_{m=c}^{k}\binom km p^m(1-p)^{k-m}},
\]
where $B_m\sim\mathrm{Bin}(m,a)$ and $B'_{k-m}\sim\mathrm{Bin}(k-m,b)$ are independent. With common parameters the number of literals equal to one at $t$ is Bin$(k,p)$. Given that $m$ of them equal one, each of these stays at one with probability $a$ and each of the other $k-m$ switches to one with probability $b$, independently, so the count at $t+1$ has the distribution of $B_m+B'_{k-m}$. The rule drops the stock when this count falls below $c$, and averaging over $m\ge c$ with weights proportional to $\binom km p^m(1-p)^{k-m}$ gives the formula. For $c=k$ it reduces to \cref{prop:gates}(i), and for $c=1$ to \cref{prop:gates}(ii).

\FloatBarrier
\section{Additional evidence on turnover and costs}\label{oa:costs}

\begin{figure}[t]
\centering
\includegraphics[width=\textwidth]{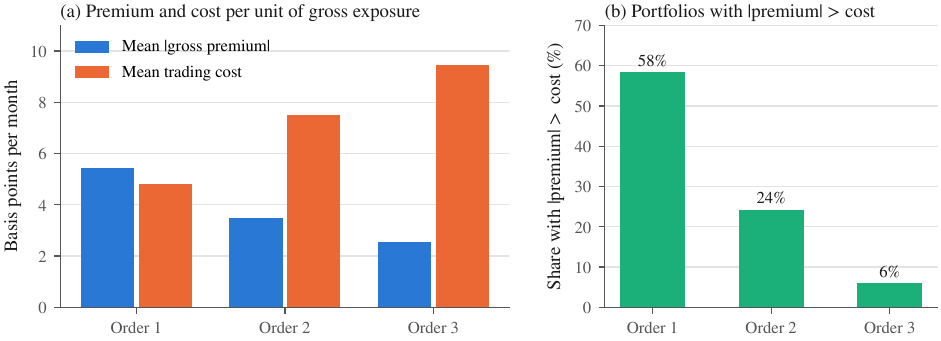}
\caption{Premia and trading costs by order, 1995--2024. Panel (a) reports the mean absolute premium and the mean quoted-spread cost of median portfolios per unit of gross exposure. Panel (b) reports the share of portfolios whose absolute premium exceeds their cost.}
\label{fig:costs}
\end{figure}

\paragraph{Turnover and costs by order.}
\Cref{tab:turnover} reports the persistence and turnover comparisons plotted in \cref{fig:turnover}, and \cref{fig:costs} plots the averages in Panel A of \cref{tab:costs}. \Cref{tab:costsens} repeats the cost comparison of \cref{tab:costs} with equal weights. Costs are higher because small stocks receive more weight, and the share of portfolios whose absolute premia exceed their costs falls from 75\% at first order to 17\% and 2\% at second and third order.

\paragraph{Dependence between signals.}
\Cref{tab:independence} groups the persistence and turnover predictions of \cref{fig:turnover} by three kinds of dependence between the signals involved: their level correlation, their flip correlation, and membership in the same economic family.

\paragraph{Other cost measures.}
Quoted spreads are simple and transparent, but small trades in liquid stocks often execute inside the quotes, and large trades move prices. Effective-spread estimates \citep{Hasbrouck2009,ChenVelikov2023} and estimates from institutional executions \citep{FIM2018} could replace them. Because the ranking of costs by order comes from differences in turnover, it is not specific to quoted spreads, and when costs are quadratic in the size of a trade, \cref{cor:profile}(ii) shows that the quadratic form that governs turnover also governs expected costs.

\begin{table}[t]\centering
\caption{Persistence of characters and turnover of AND, OR, and majority rules}\label{tab:turnover}
\begin{threeparttable}\small
\begin{tabular}{lccccc}\toprule
\multicolumn{6}{l}{\emph{Panel A. Character persistence} $\E[\chi_S(\mathbf x_t)\chi_S(\mathbf x_{t+1})]$}\\
Order & Characters & Realized & $\prod_{j\in S}\rho_j$ & Correlation & Mean abs.\ diff. \\ \midrule
1 & 12 & 0.796 & 0.796 & 1.000 & 0.000 \\
2 & 66 & 0.629 & 0.626 & 1.000 & 0.003 \\
3 & 220 & 0.493 & 0.486 & 0.999 & 0.007 \\
4 & 495 & 0.382 & 0.371 & 0.999 & 0.010 \\
\midrule
\multicolumn{6}{l}{\emph{Panel B. One-way turnover of equal-weighted long legs}}\\
Rule & $k$ & Coverage & Realized & Predicted & \\ \midrule
AND & 1 & 0.501 & 0.101 & 0.102 \\
AND & 2 & 0.249 & 0.165 & 0.163 \\
AND & 3 & 0.139 & 0.254 & 0.256 \\
AND & 4 & 0.070 & 0.421 & 0.425 \\
AND & 5 & 0.034 & 0.485 & 0.474 \\
\addlinespace
OR & 1 & 0.501 & 0.101 & 0.102 \\
OR & 2 & 0.746 & 0.071 & 0.071 \\
OR & 3 & 0.863 & 0.052 & 0.049 \\
OR & 4 & 0.939 & 0.027 & 0.029 \\
OR & 5 & 0.961 & 0.015 & 0.014 \\
\addlinespace
Majority & 1 & 0.501 & 0.101 & 0.102 \\
Majority & 3 & 0.503 & 0.150 & 0.147 \\
Majority & 5 & 0.503 & 0.140 & 0.154 \\
\bottomrule\end{tabular}
\begin{tablenotes}\footnotesize
\item Panel A compares the realized month-to-month persistence of each Walsh character of order one to four with the product of literal persistences implied by Proposition~\ref{prop:noise}. Panel B reports averages over up to 25 randomly drawn sets of $k$ literals per rule and size. Predicted turnover is the exact value under Assumption~\ref{ass:markov} at the estimated $\rho_j$, computed by enumerating the $2^k$ literal states (Proposition~\ref{prop:gates} gives the AND and OR cases in closed form). Across all 286 legs the slope of realized on predicted turnover is 1.016 and the correlation is 0.993.
\end{tablenotes}\end{threeparttable}\end{table}

\begin{table}[t]\centering
\caption{Turnover predictions and the dependence between signals}\label{tab:independence}
\begin{threeparttable}\footnotesize\setlength{\tabcolsep}{4pt}
\begin{tabular}{lcccccccc}\toprule
\multicolumn{9}{l}{\emph{Panel A. Persistence of median characters}}\\
 & \multicolumn{4}{c}{Order 2} & \multicolumn{4}{c}{Order 3} \\ \cmidrule(lr){2-5}\cmidrule(lr){6-9}
Signals involved & $N$ & Realized & Predicted & Error & $N$ & Realized & Predicted & Error \\ \midrule
\multicolumn{9}{l}{Largest flip correlation} \\
\quad below 0.05 & 60 & 0.600 & 0.599 & 0.001 & 168 & 0.435 & 0.432 & 0.003 \\
\quad 0.05 to 0.15 & 3 & 0.928 & 0.919 & 0.009 & 24 & 0.669 & 0.661 & 0.008 \\
\quad 0.15 or more & 3 & 0.914 & 0.878 & 0.035 & 28 & 0.694 & 0.664 & 0.030 \\
\addlinespace
\multicolumn{9}{l}{Largest level correlation} \\
\quad below 0.1 & 40 & 0.531 & 0.530 & 0.001 & 52 & 0.279 & 0.276 & 0.003 \\
\quad 0.1 to 0.3 & 21 & 0.772 & 0.768 & 0.004 & 120 & 0.532 & 0.528 & 0.004 \\
\quad 0.3 or more & 5 & 0.819 & 0.804 & 0.015 & 48 & 0.627 & 0.610 & 0.017 \\
\addlinespace
\multicolumn{9}{l}{Same economic family} \\
\quad no & 58 & 0.637 & 0.635 & 0.002 & 144 & 0.515 & 0.510 & 0.005 \\
\quad yes & 8 & 0.573 & 0.564 & 0.009 & 76 & 0.451 & 0.441 & 0.011 \\
\addlinespace
All & 66 & 0.629 & 0.626 & 0.003 & 220 & 0.493 & 0.486 & 0.007 \\
\bottomrule\end{tabular}

\medskip
\begin{tabular}{lccccc}\toprule
\multicolumn{6}{l}{\emph{Panel B. One-way turnover of AND, OR, and majority legs on two to five literals}}\\
Signals involved & $N$ & Realized & Predicted & Error & Relative error (\%) \\ \midrule
\multicolumn{6}{l}{Largest flip correlation} \\
\quad below 0.05 & 172 & 0.191 & 0.189 & 0.002 & 1.3 \\
\quad 0.05 to 0.15 & 26 & 0.152 & 0.153 & $-$0.001 & $-$2.8 \\
\quad 0.15 or more & 52 & 0.149 & 0.155 & $-$0.006 & $-$6.6 \\
\addlinespace
\multicolumn{6}{l}{Largest level correlation} \\
\quad below 0.1 & 47 & 0.198 & 0.194 & 0.004 & 1.5 \\
\quad 0.1 to 0.3 & 123 & 0.187 & 0.187 & 0.000 & $-$0.5 \\
\quad 0.3 or more & 80 & 0.152 & 0.155 & $-$0.003 & $-$2.6 \\
\addlinespace
\multicolumn{6}{l}{Same economic family} \\
\quad no & 122 & 0.140 & 0.139 & 0.002 & $-$1.7 \\
\quad yes & 128 & 0.214 & 0.216 & $-$0.002 & 0.1 \\
\addlinespace
All & 250 & 0.178 & 0.178 & 0.000 & $-$0.8 \\
\bottomrule\end{tabular}
\begin{tablenotes}\footnotesize
\item The flip correlation of two signals is the correlation, across stock-months, of the indicators that their median literals change between $t$ and $t+1$; the level correlation is the time-series average of the monthly cross-sectional correlation of the two literals. Each character or leg is classified by the largest absolute correlation among the pairs of signals it involves, or by whether two of its signals belong to the same economic family: size; value; past returns (momentum, short-term and long-term reversal); profitability (gross profitability, return on equity); investment and issuance (asset growth, share issuance, sales growth); and risk (volatility, beta). Predictions use \cref{prop:noise} and \cref{prop:gates} with the realized persistence of each literal. Realized, predicted, and error (realized minus predicted) are group averages; the relative error divides the error of each leg by its predicted turnover and is averaged across legs, so it can differ in sign from the average error. The largest flip correlations are 0.38 for asset growth and sales growth and 0.25 for return on equity and asset growth; value and return on equity have a level correlation of $-$0.365, and the persistence of their interaction differs from its prediction by 0.006. Across all characters or legs, the slope of realized on predicted values is 1.00 at order two, 1.01 at order three, and 1.02 for the legs.
\end{tablenotes}\end{threeparttable}\end{table}

\begin{table}[t]\centering
\caption{Premia and trading costs by order, equal weights}\label{tab:costsens}
\begin{threeparttable}\footnotesize\setlength{\tabcolsep}{4pt}
\begin{tabular}{cccccc}\toprule
Order & $N$ & Turnover & $|$Premium$|$ (bp) & Cost (bp) & $|$Premium$|>$ cost (\%) \\ \midrule
1 & 12 & 0.252 & 6.88 & 6.07 & 75.0 \\
2 & 66 & 0.441 & 3.50 & 10.04 & 16.7 \\
3 & 220 & 0.559 & 2.56 & 12.52 & 2.3 \\
\bottomrule\end{tabular}
\begin{tablenotes}\footnotesize
\item The table repeats Panel A of Table~\ref{tab:costs} with equal base weights, under which every median literal has coverage one half and each second-order portfolio is a pure interaction contrast. Median portfolios, 1995--2024; premium and cost are per unit of gross exposure and per month.
\end{tablenotes}\end{threeparttable}\end{table}

\FloatBarrier
\section{Additional evidence on pricing}\label{oa:pricing}

\begin{figure}[t]
\centering
\includegraphics[width=\textwidth]{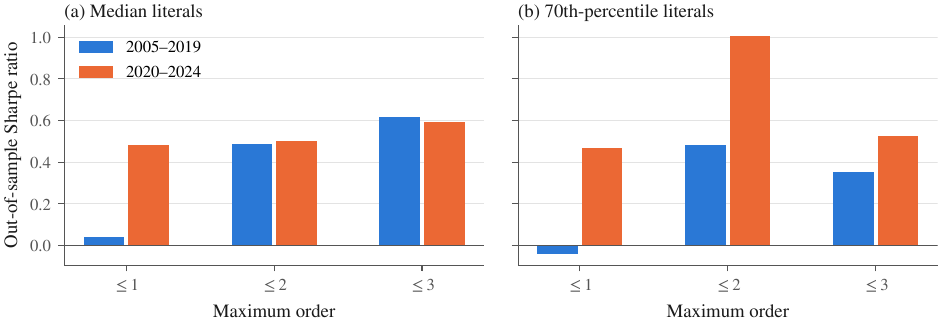}
\caption{Out-of-sample Sharpe ratios of tangency portfolios of order at most one, two, and three, over 2005--2019 and 2020--2024. Weights are ridge estimates from 1973--2004. Panel (a) uses median literals; panel (b) uses literals at the 70th percentile of each oriented signal.}
\label{fig:oosdegree}
\end{figure}

\paragraph{Order blocks.}
\Cref{tab:blocks} reports the block GRS tests and the out-of-sample comparisons of \cref{sec:info} for median literals with capped value and equal weights and for literals at the 70th percentile. \Cref{fig:oosdegree} splits the out-of-sample period. From 2005 to 2019 the first-order tangency portfolio earns a Sharpe ratio of 0.04 and the tangency portfolio of order at most two 0.48; from 2020 to 2024 both earn about 0.5.

\begin{figure}[t]
\centering
\includegraphics[width=\textwidth]{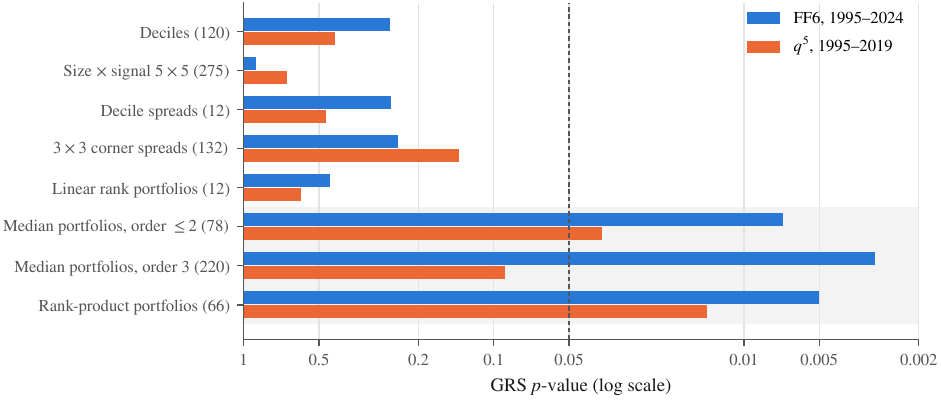}
\caption{GRS $p$-values of FF6 (1995--2024) and $q^5$ (1995--2019) for test assets built from the same twelve signals. Bars extend to the right as $p$ falls; the dashed line marks $p=0.05$. The shaded rows are median and rank-product portfolios. Numbers in parentheses are the numbers of test assets.}
\label{fig:testassets}
\end{figure}

\paragraph{Test assets.}
\Cref{fig:testassets} plots the GRS $p$-values of \cref{tab:testassets}, and \cref{tab:windows} reports the FF6 tests for samples that start in different years.

\paragraph{Missing returns.}
\Cref{tab:missing} replaces the zero excess returns of stock-months without a return for month $t+1$ by CRSP delisting returns and repeats the block tests, the out-of-sample tangency portfolios, the cost comparison by order, and the FF6 tests of the main test-asset sets. No statistic changes materially.

\begin{table}[t]\centering
\caption{The order structure of expected returns}\label{tab:blocks}
\begin{threeparttable}\footnotesize\setlength{\tabcolsep}{4pt}
\begin{tabular}{lccccccc}\toprule
 & \multicolumn{2}{c}{Block GRS $p$, 1973--2024} & \multicolumn{3}{c}{OOS SR, 2005--2024} & \multicolumn{2}{c}{OOS $t(\alpha)$, 2005--2024} \\ \cmidrule(lr){2-3}\cmidrule(lr){4-6}\cmidrule(lr){7-8}
 & order 2 on $\le1$ & order 3 on $\le2$ & $\le1$ & $\le2$ & $\le3$ & $\le2$ on $\le1$ & $\le3$ on $\le2$ \\ \midrule
Median, capped VW & $<$0.001 & 0.006 & 0.18 & 0.48 & 0.61 & 2.46 & 1.76 \\
Median, EW & $<$0.001 & 0.037 & 0.36 & 0.70 & 0.88 & 2.46 & 2.46 \\
70th percentile, capped VW & $<$0.001 & 0.276 & 0.13 & 0.64 & 0.41 & 3.27 & $-$0.28 \\
\bottomrule\end{tabular}
\begin{tablenotes}\footnotesize
\item Rows differ in the threshold of the literals and in the base weights. Block GRS tests whether portfolios of a given order have zero intercepts on all portfolios of lower order. The out-of-sample (OOS) columns use ridge tangency weights estimated on 1973--2004, with the penalty chosen by five-fold cross-validation with contiguous folds inside that window; $t(\alpha)$ is the Newey--West $t$-statistic (six lags) of the $\alpha$ of one tangency portfolio on the other. Sharpe ratios (SR) are annualized.
\end{tablenotes}\end{threeparttable}\end{table}

\begin{table}[t]\centering
\caption{FF6 GRS $p$-values for samples that start in different years}\label{tab:windows}
\begin{threeparttable}\footnotesize\setlength{\tabcolsep}{4pt}
\begin{tabular}{ccccccccc}\toprule
 & & \multicolumn{4}{c}{Conventional sorts} & \multicolumn{2}{c}{Median portfolios} & \\ \cmidrule(lr){3-6}\cmidrule(lr){7-8}
Start & $T$ & Deciles & Size $\times$ signal & Decile spreads & Corner spreads & Order $\le2$ & Order 3 & Rank products \\ \midrule
1973 & 623 & $<$0.001 & 0.001 & 0.001 & $<$0.001 & $<$0.001 & $<$0.001 & $<$0.001 \\
1980 & 540 & $<$0.001 & 0.238 & 0.034 & $<$0.001 & $<$0.001 & $<$0.001 & $<$0.001 \\
1985 & 480 & 0.001 & 0.724 & 0.046 & 0.003 & $<$0.001 & 0.002 & $<$0.001 \\
1990 & 420 & 0.029 & 0.749 & 0.217 & 0.167 & $<$0.001 & 0.019 & $<$0.001 \\
1995 & 360 & 0.260 & 0.895 & 0.258 & 0.241 & 0.007 & 0.003 & 0.005 \\
2000 & 300 & 0.274 & 0.372 & 0.431 & 0.376 & 0.010 & 0.119 & 0.060 \\
2005 & 240 & 0.673 & -- & 0.685 & 0.246 & 0.004 & 0.066 & 0.084 \\
\bottomrule\end{tabular}
\begin{tablenotes}\footnotesize
\item GRS tests of FF6 on the test-asset sets of \cref{tab:testassets}, for samples that start in January of the given year and end in December 2024; $T$ is the number of months. Capped value weights. A dash marks sets with too many portfolios relative to $T$ for the GRS test.
\end{tablenotes}\end{threeparttable}\end{table}

\begin{table}[t]\centering
\caption{Missing returns replaced by CRSP delisting returns}\label{tab:missing}
\begin{threeparttable}\footnotesize\setlength{\tabcolsep}{4pt}
\begin{tabular}{lcc}\toprule
 & Main & Delisting \\ \midrule
Block GRS $p$, order 2 on order $\le1$, 1973--2024 & $<$0.001 & $<$0.001 \\
Block GRS $p$, order 3 on order $\le2$, 1973--2024 & 0.006 & 0.005 \\
Out-of-sample Sharpe ratio, order $\le1$, gross / net & 0.18 / 0.11 & 0.18 / 0.11 \\
Out-of-sample Sharpe ratio, order $\le2$, gross / net & 0.48 / 0.32 & 0.49 / 0.33 \\
Out-of-sample Sharpe ratio, order $\le3$, gross / net & 0.61 / 0.45 & 0.61 / 0.46 \\
Out-of-sample $t(\alpha)$, order $\le2$ on order $\le1$ & 2.46 & 2.48 \\
$|$Premium$|>$ cost, order 1 (\%) & 58.3 & 58.3 \\
$|$Premium$|>$ cost, order 2 (\%) & 24.2 & 24.2 \\
$|$Premium$|>$ cost, order 3 (\%) & 5.9 & 5.9 \\
FF6 GRS $p$, median portfolios, order $\le2$ & 0.007 & 0.011 \\
FF6 GRS $p$, median portfolios, order 3 & 0.003 & 0.003 \\
FF6 GRS $p$, rank products & 0.005 & 0.006 \\
FF6 GRS $p$, deciles & 0.260 & 0.244 \\
FF6 GRS $p$, corner spreads & 0.241 & 0.225 \\
\bottomrule\end{tabular}
\begin{tablenotes}\footnotesize
\item Main is the treatment used in the paper, which assigns a zero excess return to stock-months without a return for month $t+1$. Delisting replaces these zeros by the CRSP delisting return when the stock delists in month $t+1$, which covers 83\% of them; the rest, mostly stocks that delisted during month $t$, keep a zero excess return. Every performance-related delisting among them has a CRSP delisting return, so the corrections of \citet{Shumway1997} and \citet{ShumwayWarther1999} are not needed. Capped value weights; out-of-sample statistics cover 2005--2024 with weights estimated on 1973--2004 (\cref{tab:blocks}), cost statistics 1995--2024 (\cref{tab:costs}), and FF6 tests 1995--2024 (\cref{tab:testassets}). $t(\alpha)$ uses Newey--West standard errors with six lags.
\end{tablenotes}\end{threeparttable}\end{table}

\FloatBarrier
\section{Rule selection: enumeration and permutation inference}\label{oa:search}

Because the space of rules on a fixed set of literals is finite, it can be searched exhaustively. This appendix shows why searching for the best rule is a different task from analyzing a given one: the best rules found in 1973--2004 are highly significant relative to a permutation null, but their performance deteriorates sharply over 2005--2024.

\paragraph{Enumeration.}
We enumerate every Boolean function of $k\le4$ of the twelve median literals. Counting a function and its complement once and dropping functions that ignore some of their inputs leaves 12, 330, 23,980, and 15,987,015 distinct rules for $k=1,\dots,4$. Because a rule on $k$ literals is constant within each of the $2^k$ cells, all sixteen million $t$-statistics follow directly from the cell returns. To calibrate the search, we permute next-month returns across stocks within each month of the 1973--2004 training sample and repeat the full search in each of 10,000 permutations.

\paragraph{The permutation test.}
The null hypothesis is sharp: conditional on the literals and base weights of all stocks in all months, the joint distribution of the return vectors is invariant to independent permutations of stocks within each month. Under this null the observed maximal $|t|$ and its permuted counterparts are exchangeable, so the rank of the observed value among the 10,001 values is uniformly distributed, and the $p$-value $(1+\#\{\text{permuted maxima}\ge\text{observed maximum}\})/10{,}001$ is exact. The test controls the family-wise error rate in the weak sense, that is, when no rule has a premium; strong control for individual rules requires further conditions such as subset pivotality \citep{WestfallYoung1993,RomanoWolf2005}. Because the null is sharp, it also rules out any link between the literals and the dispersion of returns, so a rejection indicates dependence between returns and literals rather than a mean premium specifically.

\begin{figure}[t]
\centering
\includegraphics[width=\textwidth]{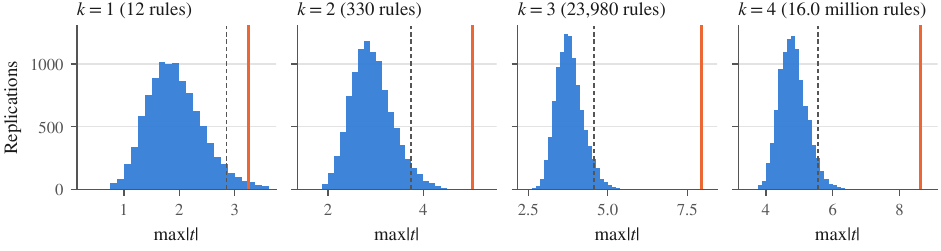}
\caption{Permutation null of the maximal $|t|$ over all rules on $k$ median literals, 1973--2004, 10,000 permutations. Each permutation shuffles next-month returns across stocks within each month and repeats the full search. Dashed lines mark the 95\% quantile; solid lines mark the best rule in the actual data.}
\label{fig:permutation}
\end{figure}

\paragraph{Results.}
\Cref{tab:search} reports the search, and \cref{fig:permutation} shows the permutation distributions. The best four-literal rule reaches $t=8.6$ in the training sample, and the best three-literal rule, an OR of book-to-market, momentum, and short-term reversal, reaches $t=7.9$. The 95\% quantiles of the permutation distribution are only 5.56 and 4.57, respectively, and in both searches none of the 10,000 permutations produces a maximum as large as the observed one ($p=1/10{,}001$). The $p$-values are 0.0002 for two-literal rules and 0.016 for single literals. Over 2005--2024, the 100 best four-literal rules keep their training sign but have a mean $t$-statistic of 1.4; for smaller $k$ the mean is between 0.17 and 1.02, and only 50\% to 86\% keep their sign.

\paragraph{Effective number of trials, deflated Sharpe ratios, and overfitting.}
The effective number of independent trials, $N_{\mathrm{eff}}=\ln0.95/\ln\bigl(2\Phi(c_{0.95})-1\bigr)$, is the number of independent standard normal statistics whose maximal absolute value has the same 95\% quantile $c_{0.95}$ as the permutation distribution. For four-literal rules it is about 1.9 million, far fewer than the sixteen million rules, because rules that share literals are strongly correlated. Because rules come in sign pairs, the deflated Sharpe ratio of \citet{BaileyLdP2014} uses $2N_{\mathrm{eff}}$ one-sided trials and the variance of Sharpe ratios across all signed rules; it lies between 0.31 and 0.80 for the best rule at each $k$, below the conventional threshold of 0.95. Its benchmark treats the whole dispersion of Sharpe ratios across rules as noise, and here that dispersion is large: the mean squared $t$-statistic across rules is about three, against one under the permutation null. The probability of backtest overfitting \citep{BBLZ2017}, computed with 16 blocks, is 0.014 for all rules on up to three literals and 0.040 for the monotone subset, whose sign is fixed by the orientation of the signals. When the sample is split into two halves, the best rule in one half almost always ranks above the median in the other, but its Sharpe ratio there is typically about 40\% lower.

\begin{table}[t]\centering
\caption{Searching the space of rules}\label{tab:search}
\begin{threeparttable}\footnotesize\setlength{\tabcolsep}{4pt}
\begin{tabular}{cccccccccc}\toprule
 & & & \multicolumn{2}{c}{Permutation null} & \multicolumn{3}{c}{Best rule} & \multicolumn{2}{c}{Top 100, 2005--2024} \\ \cmidrule(lr){4-5}\cmidrule(lr){6-8}\cmidrule(lr){9-10}
$k$ & Rules & Monotone & $q_{0.95}(\max|t|)$ & $N_{\mathrm{eff}}$ & $\max|t|$ & $p$ & DSR & Mean $t$ & Same sign (\%) \\ \midrule
1 & 12 & 12 & 2.85 & 12 & 3.25 & 0.016 & 0.36 & 0.17 & 50 \\
2 & 330 & 132 & 3.74 & 281 & 5.04 & 0.0002 & 0.31 & 0.51 & 78 \\
3 & 23{,}980 & 1{,}980 & 4.57 & 10{,}410 & 7.94 & 0.0001 & 0.80 & 1.02 & 86 \\
4 & 15{,}987{,}015 & 56{,}430 & 5.56 & 1{,}885{,}665 & 8.64 & 0.0001 & 0.48 & 1.41 & 100 \\
\bottomrule\end{tabular}
\begin{tablenotes}\footnotesize
\item Rules is the number of Boolean functions of exactly $k$ of the twelve median literals, counting a function and its complement once; Monotone is the number of those that are monotone. Statistics use 1973--2004. The permutation null permutes next-month returns across stocks within each month and repeats the full search (10{,}000 permutations). $N_{\mathrm{eff}}$ is the number of independent standard normal statistics whose maximal absolute value has the same 95\% quantile. $p$ is the permutation $p$-value of the best rule, counting the observed statistic among the 10{,}001 values. DSR is the deflated Sharpe ratio of the best rule, with $2N_{\mathrm{eff}}$ one-sided trials and the cross-rule variance of Sharpe ratios. The last two columns report the out-of-sample $t$-statistics of the 100 rules with the largest training $|t|$ (all 12 for $k=1$), signed by the training sign. The probability of backtest overfitting (16 blocks) is 0.014 for all rules with $k\le3$ and 0.040 for the monotone subset.
\end{tablenotes}\end{threeparttable}\end{table}

\FloatBarrier
\section{Construction choices as rule disagreements}\label{oa:nse}

\begin{figure}[t]
\centering
\includegraphics[width=\textwidth]{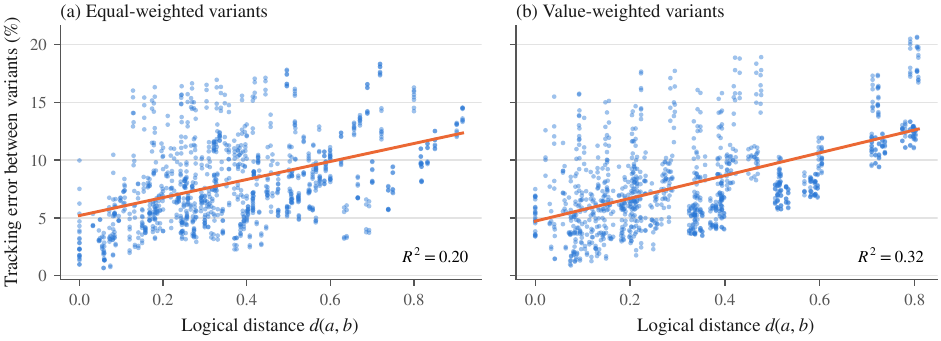}
\caption{Tracking error between pairs of construction variants of the value factor against their logical distance $d(a,b)$, 1973--2024. Panel (a) uses pairs of equal-weighted variants and panel (b) pairs of value-weighted variants.}
\label{fig:nse}
\end{figure}

The representation can also be used to audit construction choices. Studies of non-standard errors show that reasonable choices in the construction of a factor move its premium by as much as sampling error \citep{Menkveld2024,WWW2024}. In the Boolean representation, two construction variants $a$ and $b$ assign each stock a code $c^a_{i,t},c^b_{i,t}\in\{-1,0,1\}$, and their logical distance in month $t$ is $d_t(a,b)=\sum_i\omega_{i,t}|c^a_{i,t}-c^b_{i,t}|$; we write $d(a,b)$ for its average over months. We build 96 variants of the value factor from three universes, two breakpoint sets, four threshold pairs, two weighting schemes, and single or size-neutral sorts. Their Sharpe ratios range from 0.08 to 0.95 over 1973--2024. \Cref{fig:nse} shows that the logical distance between two variants explains 20\% of the variation in their tracking error when both are equal-weighted and 32\% when both are value-weighted. For two variants that share the universe and base weights, $(c^a_{i,t}-c^b_{i,t})^2\le2|c^a_{i,t}-c^b_{i,t}|$ and the Cauchy--Schwarz inequality give $|F^a_{t+1}-F^b_{t+1}|\le\bigl(2\,d_t(a,b)\sum_i\omega_{i,t}\tilde r_{i,t+1}^2\bigr)^{1/2}$ for the managed portfolios of the two codes, so logical distance bounds the difference in returns. The remaining variation comes from the returns of the stocks on which the variants disagree.

\FloatBarrier
\section{Minimization and verification of an estimated rule}\label{oa:verify}

Because a rule is a finite logical object, properties such as non-overlapping legs or required exclusions can be checked exactly for every possible stock profile. As an illustration, we estimate a mean-variance combination of the median portfolios of order at most two on 1973--2004, with a penalty proportional to the turnover $1-\rho_S$ of each portfolio, and threshold its score at the capped value-weighted median to obtain a long rule $L^{+}$ on the twelve median literals, and define a short rule $L^{-}$ from the bottom quartile. Of the 4,096 literal profiles, 2,428 enter the long leg. Profiles never observed in the training sample are left unassigned. Logic minimization reduces the rule to a disjunction of 166 conjunctions with 908 literals in total, and its reduced ordered binary decision diagram \citep{Bryant1986} has 251 nodes. A satisfiability-modulo-theories solver \citep{deMouraBjorner2008} proves that $L^{+}\wedge L^{-}$ has no satisfying profile, so the legs never overlap, and that the minimized rule agrees with the original on every observed profile. It also finds that $L^{+}$ admits stocks in the high-volatility half and the high-beta half at the same time, which an exclusion rule might forbid. The repaired rule $L^{+}\wedge\neg(\neg x_{\mathrm{VOL}}\wedge\neg x_{\mathrm{BETA}})$ meets the exclusion by construction and keeps 73\% of the long leg's capped market value.

\end{document}